\documentclass[a4paper]{article}
\usepackage[margin=1.1in]{geometry}
\usepackage{authblk}
\usepackage{setspace}
\usepackage{array}
\usepackage{makecell}
\usepackage{amsmath, amssymb, booktabs, mathtools}
\usepackage{subfig}
\usepackage[font=small,labelfont=bf]{caption}
\usepackage{threeparttable}
\usepackage{float}
\usepackage[square, sort&compress, numbers]{natbib}
\usepackage{bm}

\usepackage{graphicx} 
\usepackage{hyperref} 
\usepackage[nameinlink,capitalize]{cleveref}
\hypersetup{
  colorlinks   = true,
  urlcolor     = blue,
  linkcolor    = blue,
  citecolor    = blue
}
\DeclareMathOperator{\cumsum}{cumsum}
\DeclareMathOperator{\diag}{diag}

\newcommand{\correspondingA}{*}

\title{Physics-guided impact localisation and force estimation in composite plates with uncertainty quantification}

\author[1]{
Dong Xiao\textsuperscript{*}
}
\author[1]{
Zahra Sharif-Khodaei
}
\author[1]{
M. H. Aliabadi
}

\affil[1]{\normalsize \slshape  Department of Aeronautics, Imperial College London, South Kensington, London SW7 2AZ, United Kingdom.}
\date{}

\begin{document}
\maketitle

\footnotetext[0]{ \textsuperscript{\correspondingA}Corresponding author}
\footnotetext[1]{Email addresses: d.xiao21@imperial.ac.uk (D. Xiao); z.sharif-khodaei@imperial.ac.uk (Z. Sharif-Khodaei); m.h.aliabadi@imperial.ac.uk (M.H. Aliabadi)}
\footnotetext[2]{ ORCID: D. Xiao \href{https://orcid.org/0009-0006-7609-7832}{\includegraphics[scale=0.04]{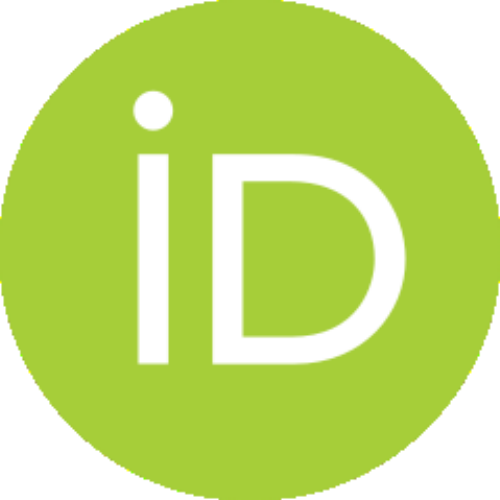}}; Z. Sharif-Khodaei \href{https://orcid.org/0000-0001-5106-2197}{\includegraphics[scale=0.04]{fig_orcidicon.pdf}} ; M.H. Aliabadi \href{https://orcid.org/0000-0002-2883-2461}{\includegraphics[scale=0.04]{fig_orcidicon.pdf}}
}

\renewcommand{\abstractname}{Abstract}
\begin{abstract}{\normalsize \onehalfspacing
Physics-guided approaches offer a promising path toward accurate and generalisable impact identification in composite structures, especially when experimental data are sparse. This paper presents a hybrid framework for impact localisation and force estimation in composite plates, combining a data-driven implementation of First-Order Shear Deformation Theory (FSDT) with machine learning and uncertainty quantification. The structural configuration and material properties are inferred from dispersion relations, while boundary conditions are identified via modal characteristics to construct a low-fidelity but physically consistent FSDT model. This model enables physics-informed data augmentation for extrapolative localisation using supervised learning. Simultaneously, an adaptive regularisation scheme derived from the same model improves the robustness of impact force reconstruction. The framework also accounts for uncertainty by propagating localisation uncertainty through the force estimation process, producing probabilistic outputs. Validation on composite plate experiments confirms the framework’s accuracy, robustness, and efficiency in reducing dependence on large training datasets. The proposed method offers a scalable and transferable solution for impact monitoring and structural health management in composite aerostructures.}
\end{abstract}

\renewcommand{\abstractname}{Keywords}
\begin{abstract}{
Structural health monitoring; Impact identification; Data-driven FSDT modelling; Physics-augmented machine learning; Physics-adaptive regularisation; Uncertainty quantification}
\end{abstract}

\newpage

\section{Introduction}
Composite aerostructures in service are frequently exposed to various impact threats, such as tool drops, hail, and bird strikes \cite{vincent_damage_2011}. Among these, low-velocity impacts pose a particular risk due to their ability to induce barely visible impact damage (BVID), which may degrade structural integrity without obvious surface indications. Structural Health Monitoring (SHM) systems, especially those employing passive sensing strategies, are widely used to detect such events. By capturing the dynamic response of the structure through surface-mounted or embedded sensors, these systems aim to infer the impact location, force history, and potential damage. However, accurately identifying impact characteristics from sparse sensor measurements remains a significant challenge, especially for high-energy impacts that lead to BVID \cite{aliabadi_structural_2018}.

Impact identification is inherently an inverse problem, wherein unknown impact parameters—location and force—must be inferred from observed structural responses. As summarised in \cref{TB: Impact identification methodologies}, existing approaches are broadly classified into model-based and data-driven methods.

\begin{table}[ht]
\centering
\caption{Impact identification methodologies.} 
\label{TB: Impact identification methodologies}
\renewcommand{\arraystretch}{1.15} 
\begin{tabular}{lllll}
\toprule
Method & Unknowns & Observations & \begin{tabular}{@{}c@{}} Required \\ knowledge \end{tabular}  & Models / Surrogates  \\
\midrule
Model-based & $q, f(t)$ & \begin{tabular}{@{}c@{}}  Impact sensor \\ signals $s(t)$ \end{tabular}  &  \begin{tabular}{@{}l@{}}Model $\mathcal{M},$ \\ Sensors $L$ \end{tabular}  & \begin{tabular}{@{}l@{}} Finite element analysis \cite{hu_efficient_2007, xiao_impact_2024, liu_quantification_2023, yu_impact_2023, yu_accelerated_2024} \\ Modal analysis \cite{goutaudier_impulse_2020, goutaudier_single-sensor_2020, zhang_efficient_2024} \\  Transfer function \cite{el-bakari_assessing_2014, atobe_impact_2014, ciampa_impact_2012, simone_hierarchical_2019, chen_impact_2012, liu_novel_2018} \\ Dispersion relations \cite{kundu_acoustic_2014, meo_impact_2005, dehghan_niri_nonlinear_2014, niri_probabilistic_2012, zhu_two-step_2018, zhu_passive_2020, ciampa_acoustic_2010, deng_multi-frequency_2024, kirkby_impact_2011, sikdar_low-velocity_2022} \end{tabular} \\ 

Data-driven & $q, f(t)$ & \begin{tabular}{@{}l@{}} Impact sensor \\ signals $s(t)$ \end{tabular}  & \begin{tabular}{@{}l@{}} Reference \\ impact  data  \\ $Q, F(t), S(t)$ \end{tabular}  & \begin{tabular}{@{}l@{}} Multi-layer perceptrons \cite{seno_passive_2019, seno_impact_2019, sharif-khodaei_determination_2012, ghajari_identification_2013, hossain_inverse_2018, morse_reliability_2018, jang_impact_2019, balasubramanian_comparison_2023} \\  Gaussian process regression \cite{xiao_robust_2025, seno_uncertainty_2021, jones_bayesian_2022, ojha_probabilistic_2024, hensman_locating_2010} \\ Convolutional neural networks \cite{tabian_convolutional_2019, zhao_impact_2024, hamadneh_impact-force_2024, hesser_identification_2021, zhou_impact_2024} \\ Recurrent neural networks \cite{huang_hybrid_2023, zhou_data-physics_2024} \\ Graph neural networks \cite{zhao_spatial-temporal_2023, huang_impact_2023} \end{tabular} \\
\bottomrule
\end{tabular}\\
    \begin{tablenotes}    
      \footnotesize               
      \item[1] *$q, f(t)$: impact location and force of the target impact (to be estimated), $\mathcal{M}$: model, $L$: Sensors' locations, $Q, F(t), S(t)$: reference impact locations, forces and sensor signals.
    \end{tablenotes}            
\end{table}

Model-based techniques rely on physical representations of structural dynamics, including finite element analysis (FEA) \cite{hu_efficient_2007, xiao_impact_2024, liu_quantification_2023, yu_impact_2023, yu_accelerated_2024}, modal analysis \cite{goutaudier_impulse_2020, goutaudier_single-sensor_2020, zhang_efficient_2024}, and transfer function formulations \cite{el-bakari_assessing_2014, atobe_impact_2014, ciampa_impact_2012, simone_hierarchical_2019, chen_impact_2012, liu_novel_2018}. Dispersion relation-based methods \cite{kundu_acoustic_2014, meo_impact_2005, dehghan_niri_nonlinear_2014, niri_probabilistic_2012, zhu_two-step_2018, zhu_passive_2020, ciampa_acoustic_2010, deng_multi-frequency_2024, kirkby_impact_2011, sikdar_low-velocity_2022}, which capture frequency-dependent wave propagation characteristics, have also been employed to enhance localisation. While these methods offer high physical fidelity, they are often limited by the availability of accurate structural information, including geometric configurations, material properties, and boundary conditions \cite{szabo_finite_2021, ereiz_review_2022}. Incomplete or inaccurate models degrade identification accuracy, and full-scale simulations typically require costly model validation and updating efforts \cite{schwarz_experimental_1999, meo_impact_2005}.

In contrast, data-driven approaches construct surrogate models that learn mappings from sensor signals to impact parameters based on reference datasets. These include traditional machine learning models such as multilayer perceptrons (MLPs) \cite{seno_passive_2019, seno_impact_2019, sharif-khodaei_determination_2012, ghajari_identification_2013, hossain_inverse_2018, morse_reliability_2018, jang_impact_2019, balasubramanian_comparison_2023}, probabilistic frameworks like Gaussian Process Regression (GPR) \cite{xiao_robust_2025, hami_seno_multifidelity_2023, seno_uncertainty_2021, jones_bayesian_2022, ojha_probabilistic_2024, hensman_locating_2010}, and more recent developments in deep learning, including convolutional (CNN) \cite{tabian_convolutional_2019, zhao_impact_2024, hamadneh_impact-force_2024, hesser_identification_2021, zhou_impact_2024}, recurrent (RNN) \cite{huang_hybrid_2023, zhou_data-physics_2024}, and graph neural networks (GNN) \cite{zhao_spatial-temporal_2023, huang_impact_2023}. Despite their growing popularity, these methods are hindered by several key limitations:
\begin{itemize}
    \item Training data limitations. Numerical data can be inexpensive but may lack fidelity, whereas experimental data are high-fidelity but costly to acquire. Effective data-driven models require extensive impact testing \cite{zhou_impact_2024, zhao_impact_2024}, increasing the cost of deployment.
    \item Generalisability. Training data cannot feasibly cover all possible impact scenarios. Consequently, data-driven models struggle to generalise to untrained impact area \cite{xiao_hybrid_2024} and impact conditions, where impact force and location may be unknown. 
    \item Interpretability. Deep learning models often function as "black boxes", making it difficult to extract physically meaningful insights, limiting their acceptance in safety-critical applications.
    \item Scalability. Data-driven approaches typically require the collection of new data and retraining of the model when applied to different structures, sensor configurations, or structural complexities, reducing their adaptability.
\end{itemize}

To address the challenges associated with impact identification in composite aerostructures—particularly those related to data scarcity—a hybrid paradigm that integrates physical modelling with data-driven learning is gaining attraction. Rather than relying solely on black-box regression from sensor signals to impact characteristics, this paradigm seeks to approximate the underlying physics of impact dynamics using interpretable and generalisable models. For example, wave dispersion relations can significantly improve localisation by characterising the frequency-dependent propagation of guided waves, while force reconstruction can benefit from damping-informed models that capture the energy dissipation inherent in real structures.

This paper presents a physics-augmented framework for impact localisation and force identification that leverages a data-driven First-Order Shear Deformation Theory (FSDT) plate model. The model is constructed using only observed structural responses, without requiring prior knowledge of geometry, material properties, or boundary conditions. Dispersion curve fitting is employed to estimate effective material properties, while boundary conditions are inferred through modal analysis. The resulting FSDT model is physically interpretable and structurally consistent, and it serves as a foundation for two key components: (1) the generation of physics-consistent augmented data to improve the generalisation of impact localisation models, and (2) the formulation of an adaptive regularisation scheme for impact force deconvolution, informed by the system's modal characteristics.

Additionally, the proposed framework incorporates uncertainty quantification by propagating localisation errors into the force reconstruction process, enabling probabilistic estimations with associated confidence bounds. This capability enhances the reliability of predictions in real-world applications, particularly under conditions of limited sensor coverage or unseen impact scenarios.

Experimental validation is conducted on a composite plate subjected to controlled low-velocity impact events. Results demonstrate that the proposed methodology significantly improves robustness, scalability, and interpretability compared to conventional data-driven approaches, while also reducing the need for exhaustive training datasets. The framework is well-suited for deployment in operational composite structures where prior model information may be incomplete or unavailable.

The key contributions of this study are as follows: 
\begin{itemize} 
\item \textbf{Data-driven FSDT modelling:} A method for constructing a FSDT plate model from observed data, with material and boundary conditions inferred via dispersion and modal analyses, respectively. 
\item \textbf{Physics-augmented machine learning:} Use of the identified physical model to generate augmented training data, enabling impact localization models to generalise beyond the training domain. 
\item \textbf{Adaptive regularisation:} A novel, physics-informed regularisation approach for impact force deconvolution, that adaptively applies penalisation to stabilise inverse force estimation.
\item \textbf{Uncertainty quantification:} Integration of probabilistic modelling to propagate localisation uncertainty into force estimates, yielding confidence-aware predictions suitable for safety-critical applications. 
\end{itemize}

The remainder of this paper is organised as follows: \cref{section FSDT modelling} describes the proposed data-driven FSDT modelling framework, including physics-augmented impact localisation and physics-adaptive regularisation for force reconstruction. Experimental validation using low-velocity impact testing is detailed in \cref{section exprimental validation}. The corresponding results for FSDT model construction, impact localisation, force reconstruction and uncertainty quantification are presented in \cref{section Results and Analysis}. Finally, the main conclusions and potential directions for future research are discussed in \cref{section Conclusions and future work}.

\section{Physics-guided impact identification via sparse data-driven FSDT modelling} \label{section FSDT modelling}
In practical applications, plate-like carbon fiber-reinforced polymer (CFRP) laminated composite structures—such as flat plates, stiffened panels, and sandwich configurations—are extensively used in aerospace and related industries. These structures are therefore the primary focus of this chapter. The geometric dimensions of such components, specifically the length $a$ and width $b$, are typically determined during the design phase to satisfy manufacturing constraints and ensure effective sensor deployment. Sensor locations, denoted by $L$, are generally fixed based on these design considerations.

However, in more complex configurations such as stiffened plates and sandwich structures, the plate thickness $d$ exhibits spatial non-uniformity. For instance, stiffened regions may consist of increased local thickness $d_s$, compared to the nominal plate thickness $d_p$. Similarly, sandwich structures often integrate aluminium cores with composite laminate facesheets, introducing additional heterogeneity. In such cases, the effective material density $\rho$ is approximated as a volume-weighted average of the constituent materials to account for spatial variation in structural properties.

Given these conditions, the impact identification task within the proposed data-driven framework is cast as a probabilistic inference problem:
\begin{equation}
\begin{aligned} 
P\left(\Theta, q, f(t)| s(t), Q, F(t), S(t), L, a, b \right),
\end{aligned}
\end{equation}
where the plate dimensions $(a, b)$, sensor locations $L$, reference impact data $(Q, F(t), S(t))$, and measured target responses $s(t)$ are treated as observed inputs. In contrast, uncertain parameters such as the plate thickness $d \in [d_{\mathrm{lb}}, d_{\mathrm{ub}}]$ and effective density $\rho \in [\rho_{\mathrm{lb}}, \rho_{\mathrm{ub}}]$ are treated as bounded equivalents to accommodate material and geometric variability. The goal is to infer the unknown structural parameters $\Theta$ and the target impact characteristics $(q, f(t))$ from the observed data.

\begin{figure}[htb] 
	\centering
		\includegraphics[width=1\columnwidth]{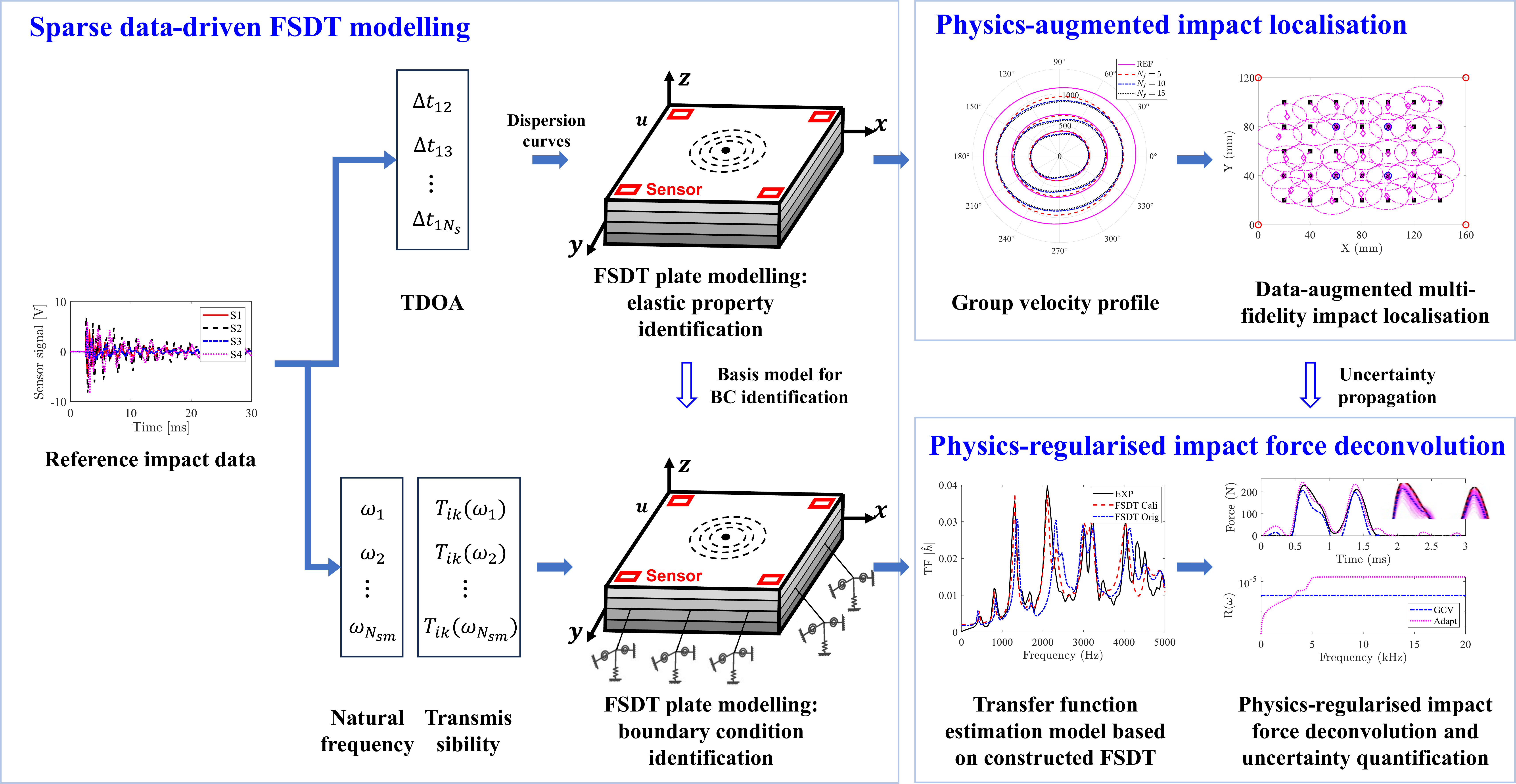}
        \captionsetup{justification=centering}
	\caption{Physics-augmented impact identification and uncertainty quantification based on sparse data-driven FSDT modelling.}
	\label{FIG: method_illustration}
\end{figure}

To enhance the generalisability of impact identification across various types of plate-like CFRP structures, a physics-based model is integrated into the framework. The FSDT is adopted to approximate the behaviour of the target structure as an equivalent flat plate. Despite geometric complexities, such as stiffeners or sandwich cores, FSDT provides a tractable yet sufficiently accurate representation of wave propagation and structural dynamics for these classes of composite systems \cite{correas_analytical_2021}. This model forms the foundation for coupling physics-based insights with data-driven techniques, thereby enabling robust physics-augmented impact localisation and physics-informed force reconstruction, as illustrated in \cref{FIG: method_illustration}. For completeness, the equations of motion governing symmetrically laminated FSDT plates are provided in \cref{Appendix A}, with detailed derivations available in standard references \cite{abrate_impact_1998, reddy_mechanics_2003}.

The construction of the FSDT-based physics model is conducted in two sequential phases:
\begin{enumerate}
    \item \textbf{Material property identification}, achieved by fitting the dispersion curve of the A0 mode;
    \item \textbf{Boundary condition calibration}, performed by matching modal characteristics such as natural frequencies and transmissibility.
\end{enumerate}
These features are extracted from the reference dataset $(Q, F(t), S(t))$, which provides time difference of arrival (TDOA) data for dispersion analysis and modal information for boundary condition tuning.

Once constructed, the FSDT model enables physics-informed impact identification through two integrated pathways:
\begin{enumerate}
    \item \textbf{Physics-guided impact localisation}: 
    Dispersion analysis $\to$ Group velocity profile (GVP) $\to$ Synthetic data generation $\to$ Data-augmented impact localisation.

    \item \textbf{Physics-guided impact force estimation}: 
    Modal analysis $\to$ Transfer function estimation $\to$ Frequency-dependent adaptive regularisation $\to$ Force deconvolution.
\end{enumerate}

For impact localisation, the estimated GVP is used to synthesise propagation-aware datasets, enabling a data-augmented multi-fidelity learning framework that enhances generalisability across the structural domain—even in regions without direct reference impacts.

Simultaneously, the FSDT model serves as a forward solver for transfer function prediction, facilitating uncertainty quantification in the radial basis function (RBF) interpolation of transfer functions \cite{simone_hierarchical_2019}. These quantified interpolation errors inform a frequency-adaptive regularisation strategy, which significantly improves the accuracy of impact force reconstruction when compared to conventional $\ell_2$ regularisation \cite{martin_impact_1996, thiene_effects_2014} with a fixed penalty parameter.

\subsection{FSDT modelling phase 1: elastic property identification}
\subsubsection{Maximum likelihood estimation of elastic properties}
The dispersion relations of a structure are inherently governed by its elastic properties while remaining independent of its boundary conditions. This property enables a hierarchical approach to identifying both structural elastic properties and boundary conditions using sparse reference impact data. Given that low-velocity impacts predominantly generate flexural waves \cite{daniel_wave_1979, tan_wave_1982}, this study focuses specifically on the analysis of out-of-plane flexural wave propagation. For symmetrically laminated composites, the dispersion relation of flexural wave group velocity is formulated as a function of the structural stiffness parameters $\boldsymbol{D}, \boldsymbol{A}_{s}$ and inertia terms $I_1, I_3$, as detailed derived in \cref{Appendix B}:
\begin{equation}
\begin{aligned} \label{dispersion relations}
v_g \left(\omega, \theta| \Theta_{p}: \{ \boldsymbol{D}, \boldsymbol{A}_{s}, I_1, I_3\} \right), \\
\end{aligned}
\end{equation} 
where $v_g(\omega, \theta)$ represents the group velocity of out-of-plane flexural waves, dependent on wave frequency $\omega$ and wave direction $\theta$ due to wave dispersion and structural anistropy. The bending stiffness matrix $\boldsymbol{D}$, share stiffness matrix $\boldsymbol{A}_{s}$ and the inertia terms $I_1, I_3$ are detailed in \cite{abrate_impact_1998, moon_theoretical_1973, tang_lamb_1988, chow_propagation_1971}.  

The group velocity profile (GVP) $v_g$ is governed by the plate-level material properties $\Theta_p$, which can be further derived from the lamina-level properties $\Theta_l$. For most-widely used transversely isotropic lamina, these lamina-level properties include the elastic constants of individual plies $E_1, E_2, G_{12}, \nu_{12}, \nu_{23}$, the lamina thickness $t$ and the stacking sequence $\Psi$:  
\begin{equation}
\begin{aligned} \label{material properties}
\Theta_l = \{ \rho, \Psi, t, E_1, E_2, G_{12}, \nu_{12}, \nu_{23} \} \to  \Theta_p = \{\boldsymbol{D}, \boldsymbol{A}_{s}, I_1, I_3 \}. 
\end{aligned}
\end{equation} 
Here, $\Psi$ is a vector of $2N_l$ elements that define the laminate stacking configuration, while the total plate thickness is given by $2N_l t_{ply}$. 

The relationship between reference impact data and wave dispersion is established through time difference of arrival (TDOA) measurements \cite{xiao_time_2024} as follows:
\begin{equation}
\begin{aligned}
\Delta t_{ij} (\omega) = t_j(\omega) - t_i(\omega) + e(\omega) = \frac{\sqrt{(x-x_j)^2+(y-y_j)^2}}{v_g(\theta_j|\omega, \Theta_l)}-\frac{\sqrt{(x-x_i)^2+(y-y_i)^2}}{v_g(\theta_i|\omega, \Theta_l)} + e(\omega),
\label{EQU C7: wave propagation for impact localisation}
\end{aligned}
\end{equation}
where $\Delta t_{1j}(\omega)$ represents the extracted TDOA between the $j$-th sensor and the $i$-th sensor, and $v_g(\theta_j|\omega, \Theta_l)$ denote the wave propagation group velocity at frequency $\omega$ from the impact point to the $j$-th sensor. The uncertainties in TDOA, denoted $e(\omega)$, are random variables influenced by wave frequency $\omega$, typically modeled as Gaussian with zero mean and variance $\sigma^2(\omega)$ \cite{dehghan_niri_nonlinear_2014}.

Since plate-level properties $\Theta_p$ primarily capture out-of-plane wave behaviour and do not account for in-plane dynamics, a more comprehensive identification of lamina-level properties $\Theta_l$ is necessary. Given the sensor location $L_j=(x_j, y_j)$, independent reference impact locations $Q_i = (x_i, y_i)$ and TDOA estimates $\Delta t^i(\omega)$ from sensor measurements, the material properties at the lamina level can be estimated using a multi-frequency maximum log-likelihood approach \cite{xiao_general_2025}:
\begin{equation}
\begin{aligned} \label{EQU C7: material properties optimisation} 
&\bar{\Theta}_l 
= \underset{\Theta_l}{\arg\max} \; \log \mathcal{L} 
= \underset{\Theta_l}{\arg\max} \; \sum_{m=1}^{N_f} \log \mathcal{L}_m
= \underset{\Theta_l}{\arg\max} \; \sum_{m=1}^{N_f} \sum_{i=1}^{N_i} \sum_{j=2}^{N_s} \log \zeta\left(\frac{e^i_{1j}(\omega_m)}{\bar{\sigma}(\omega_m)}\right), \\
&\bar{\sigma}^2(\omega_m) = \frac{1}{N_i(N_s-1)} \sum_{i=1}^{N_i} \sum_{j=2}^{N_s} \left[e^i_{1j}(\omega_m)\right]^2, \;
\mathcal{L}_m = \prod_{i=1}^{N_i} \prod_{j=2}^{N_s} \zeta\left(\frac{e^i_{1j}(\omega_m)}{\bar{\sigma}(\omega_m)}\right), 
\end{aligned}
\end{equation} 
where $\zeta$ denotes the probability density function of a standard Gaussian distribution, and $\hat{\sigma}^2(\omega_m)$ is the estimated variance of TDOA at frequency $\omega_m$, obtained by maximising the likelihood function. The likelihood function $\mathcal{L}_m$ at each frequency $\omega_m$ is derived based on the wave propagation formulation in \cref{EQU C7: wave propagation for impact localisation}. 

This multi-frequency identification approach leverages the full dispersion curve, reducing dependency on extensive reference datasets and large sensor arrays. Consequently, it enhances the feasibility of sparse data-driven applications in SHM by improving the accuracy and robustness of material property estimation while minimising experimental overhead.

\subsubsection{Design space of elastic properties of CFRP lamina}
In the maximum likelihood estimation of the elastic properties of CFRP lamina based on dispersion relations, the definition of the probability spaces for these properties is crucial. CFRP composite lamina are anisotropic, meaning their mechanical properties vary depending on the direction relative to the fibre orientation. The longitudinal modulus, $E_1$ (along the fibre direction), is significantly higher than the transverse modulus, $E_2$ (perpendicular to the fibres). For unidirectional CFRP lamina, $E_1$ typically ranges from 120 to 250 GPa \cite{e_c_s_s_structural_2011}, encompassing low-modulus lamina such as T800H fibre/epoxy \cite{e_c_s_s_structural_2011}, high-modulus lamina such as GY‐70/epoxy \cite{e_c_s_s_structural_2011}, and high-tenacity lamina such as T300/T400/T700S fibres/epoxy \cite{e_c_s_s_structural_2011} and AS4/AS6 fibres/epoxy \cite{soden_lamina_2004}.

The ratio $E_2/E_1$, which depends on both the fibre and matrix properties, typically ranges from 0.05 to 0.1 for CFRP \cite{trauth_mechanical_2016}. Another key ratio, $G_{12}/E_1$, representing the in-plane shear modulus $G_{12}$, generally falls between 0.02 and 0.05 due to fibre-matrix interactions. These ranges align with values observed in widely used unidirectional lamina \cite{e_c_s_s_structural_2011}. Collectively, these bounds define the probability space for the elastic properties of CFRP lamina, summarised in \cref{TB C7: probability space of the elastic properties}.

\begin{table}[ht]
\centering
\caption{Probability space of the elastic properties of CFRP lamina and stacking sequence} \label{TB C7: probability space of the elastic properties}
\small{
\begin{tabular}{ccccccccc}
\toprule
Properties & $E_1$ (GPa)  & $E_2$ (GPa) & $G_{12}$ (GPa) & $\nu_{12}(\nu_{23})$ & $\rho$ ($\mathrm{kg/m^3}$) & $t_{ply}$ (mm) & $\psi_k$ ($^\circ$) & $d$ (mm) \\
\midrule
UB & 250 & 25 & 12.5 & 0.35 & 1700 & 0.3 & \{45, 90\} & $d_{ub}$  \\
LB & 120 & 6 & 2.4 & 0.25 & 1500 & 0.1 & \{-45, 0\} & $d_{lb}$   \\
\bottomrule
\end{tabular}}\\
    \begin{tablenotes}    
        \footnotesize               
        \item[1] \hspace{2mm} *Note: UB: upper bound, LB: lower bound $E_1$: longitudinal modulus along the fibre direction, $E_2$: transverse modulus perpendicular to the fibre direction, $G_{12}$: in-plane shear modulus, $\nu_{12}$: in-plane Poisson's ratio, $\rho$:  material density, $d$: plate equivalent thickness, $N_l$: half of lamina number, $t_{ply}$: thickness of lamina.  
  \end{tablenotes}            
\end{table}

The in-plane Poisson’s ratio, $\nu_{12}$, represents the ratio of transverse to longitudinal strain under uniaxial loading along the fibre direction. While carbon fibres themselves exhibit a low Poisson’s ratio ($\nu_{fibre} \approx 0.2-0.3$, table 4.4-4 in \cite{e_c_s_s_structural_2011}), their influence on $\nu_{12}$ is minimal, as the transverse response is primarily governed by the matrix. The epoxy matrix, which deforms more in the transverse direction, typically has a Poisson’s ratio of $\nu_{matrix} \approx 0.3-0.4$ (table 3.2-6 in \cite{e_c_s_s_structural_2011}). For aerospace-grade CFRP (fibre volume fraction 60$\%$-70$\%$, table 4.4-5 in \cite{e_c_s_s_structural_2011}) and commercial-grade CFRP (50$\%$-60$\%$), $\nu_{12}$ generally falls between 0.25 and 0.35.

The density of CFRP lamina or laminate is determined by the volume-weighted average of fibre and matrix densities. Due to the higher fraction of fibres, they predominantly influence the composite density. Standard carbon fibres and high-modulus carbon fibres have densities of 1750-1900 $\mathrm{Kg/m^3}$ and 1900-2000 $\mathrm{Kg/m^3}$, respectively (table 3.3-1 in \cite{e_c_s_s_structural_2011}), while commonly used epoxy resins have lower densities, typically between 1100 and 1300 $\mathrm{Kg/m^3}$ (table 3.2-7 in \cite{e_c_s_s_structural_2011}). Consequently, for fibre volume fractions of 50$\%$-70$\%$, the overall density of CFRP lamina typically ranges from 1500 to 1700 $\mathrm{Kg/m^3}$.

\begin{figure}[htb] 
	\centering
		\includegraphics[width=0.7\columnwidth]{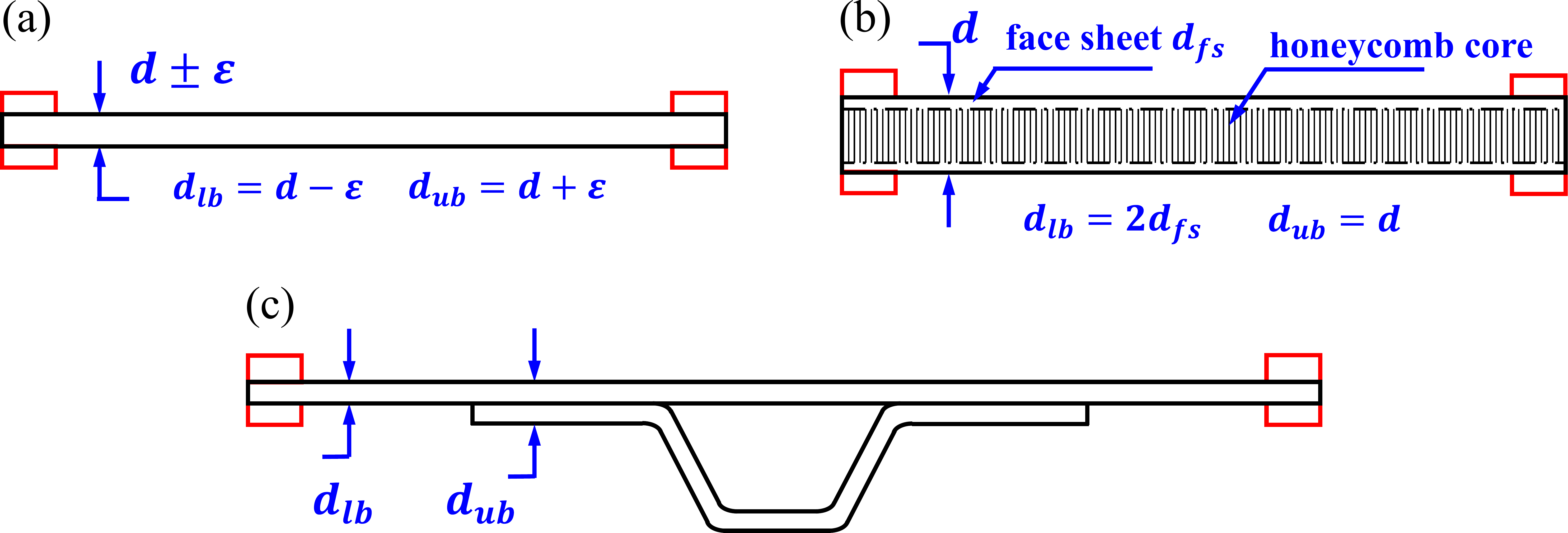}
	\caption{Three type of laminated composite plates with bounded thickness: (a) flat plate, (b) sandwich plate, (c) stringer-stiffened plate. }
	\label{FIG C7: fig_three_panels_thickness}
\end{figure}

The equivalent plate thickness is bounded based on the plate type. As illustrated in \cref{FIG C7: fig_three_panels_thickness}, for a flat plate of thickness $d$, the equivalent thickness is constrained within $[d-\varepsilon, d+\varepsilon]$, where $\varepsilon$ represents measurement uncertainty. For a sandwich plate of total thickness $d$ and face sheet thickness $d_{fs}$, the equivalent thickness falls within $[2d_{fs}, d]$. For a stringer-stiffened plate, the equivalent thickness is naturally bounded between the thickness of the unstiffened region and that of the stringer toes.

The stacking sequence of the lamina is a key factor in defining the structural anisotropy and mechanical performance of composite laminates. In this study, four of the most commonly used ply orientations in practical applications are considered: $0^\circ$, $45^\circ$, $-45^\circ$, and $90^\circ$. This selection simplifies the design space of the stacking sequence vector $\Psi$ while maintaining practical relevance. Variations in fibre orientation across individual plies influence both the in-plane and out-of-plane stiffness of the laminate, thereby altering its dispersion relations. The defined parameter spaces for material properties and lamina layup provide a structured basis for the data-driven identification of elastic properties.

The material property identification process, as formulated in \cref{EQU C7: material properties optimisation}, is a mixed-integer optimisation problem, where the four candidate ply angles are represented as discrete integer values ranging from 1 to 4. Given that the fitness function, defined as the log-likelihood, can be evaluated at relatively low computational cost—primarily involving calculations of location-to-sensor distances and wave propagation velocities—a mixed-integer genetic algorithm \cite{deep_real_2009} is employed to efficiently solve this optimisation problem and determine the optimal material properties.

\subsection{FSDT modelling phase 2: boundary condition identification} 
The identification of the elastic properties of laminated composites facilitates the subsequent determination of boundary conditions based on modal characteristics.

\subsubsection{Forward modal analysis by Rayleigh-Ritz variational method with general boundary conditions}
Based on the constitutive equations of FSDT plate, the strain energy $U$, kinetic energy $T$, and potential energy $V$ of external forces for a symmetric plate subjected to a transverse impact load $f(t)$ at a point location $q$ can be expressed as:
\begin{equation}
\begin{aligned} \label{EQU C7: strain and kinetic energy}
&U = \frac{1}{2}\int_{\Omega}[\boldsymbol{\epsilon}_0^T \boldsymbol{A} \boldsymbol{\epsilon}_0 + \boldsymbol{\kappa}^T \boldsymbol{D} \boldsymbol{\kappa} + \boldsymbol{\gamma}^T \boldsymbol{A}_s \boldsymbol{\gamma} ]d\Omega, \\
&T = \frac{1}{2}\int_{\Omega}[I_1 \dot{w}_0^2 + I_3(\dot{\varphi}_x^2+\dot{\varphi}_y^2)] d\Omega, \\ 
&V = -\int_{\Omega } f w_{0}(q)  d\Omega,
\end{aligned}
\end{equation}
where $\Omega$ represents the structural spatial domain in the $x$-$y$ plane, $\boldsymbol{\epsilon}_0$, $\boldsymbol{\kappa}$ and $\boldsymbol{\gamma}$ are vectors of midplane strains, plate curvatures, and shear strains, respectively. The boundary conditions also store energy during impact events, depending on their nature. In real-world structures, boundary conditions are rarely purely free, simply supported, or fully clamped. To model general boundary conditions, artificial springs are introduced to simulate the shear force $V_s$ and bending moment $M_b$ acting on the boundary $\Gamma$. This is expressed in terms of the spring stiffness as follows:
\begin{equation}
\begin{aligned}
V_s = k_t \mathbf{d}|_\Gamma, \; 
M_b = k_r \frac{\partial \mathbf{d}}{\partial \overrightarrow{n}}|_\Gamma
\end{aligned}
\end{equation}
where $k_t$ and $k_r$ denote the translational and rotational stiffness, respectively. Here, $\mathbf{d} = [u_0, v_0, w_0, \varphi_x, \varphi_y]^T$ represents the displacements resisted by the springs, and $\overrightarrow{n}$ is the tangential direction along the boundary. Consequently, the additional strain energy stored in the elastic boundary is given by:
\begin{equation}
\begin{aligned} \label{EQU C7: strain energy in BC}
U_b = \frac{1}{2} \int_{\Gamma} k_t \mathbf{d}^2 d\Gamma + \frac{1}{2} \int_{\Gamma} k_r(\frac{\partial \mathbf{d}}{\partial \overrightarrow{n}})^2 d\Gamma
\end{aligned}
\end{equation}
Using the Rayleigh-Ritz variational approximation approach, the displacements are assumed to take the form:
\begin{equation}
\begin{aligned} \label{EQU C7: displacement field}
&\mathbf{d} = [u_0, v_0, w_0, \varphi_x, \varphi_y]^T  = [\mathbf{U_0}\mathbf{B}^T, \mathbf{V_0}\mathbf{B}^T, \mathbf{W_0}\mathbf{B}^T, \mathbf{\Phi_{x}}\mathbf{B}^T, \mathbf{\Phi_{y}}\mathbf{B}^T]^T =  \mathbf{D} \otimes \mathbf{B}^T, \\
&\mathbf{D}= [\mathbf{U_0}, \mathbf{V_0}, \mathbf{W_0}, \mathbf{\Psi_{x}}, \mathbf{\Psi_{y}}]^T, \mathbf{B} = [b_1(\frac{x}{a})b_1(\frac{y}{b}), b_1(\frac{x}{a})b_2(\frac{y}{b}), ..., b_M(\frac{x}{a})b_N(\frac{y}{b})], 
\end{aligned}
\end{equation}
where the basis function $b_j(x)$ is chosen as Legendre polynomials due to their favourable convergence properties \cite{ni_aeroelastic_2023}, and $\mathbf{D}$ represents the coefficients of these basis functions. The operator $\otimes$ denotes the block Kronecker product. Substituting \cref{EQU C7: displacement field} into \cref{EQU C7: strain and kinetic energy} and \cref{EQU C7: strain energy in BC}, and applying Hamilton’s principle, the equations of motion are obtained as:
\begin{equation}
\begin{aligned} \label{EQU C7: Rayleigh-Ritz}
\mathbf{M} \ddot{\mathbf{D}} + (\mathbf{K}_p+\mathbf{K}_b) \mathbf{D}= \mathbf{F}, \\
\end{aligned}
\end{equation}
where $\mathbf{M}$ and $\mathbf{F}$ denote the generalised mass matrix and force vector, respectively. $\mathbf{K}_p$, $\mathbf{K}_b$ are the stiffness matrixes for the plate structure, boundary conditions, respectively. The expressions of these matrices are detailed in \cref{Appendix C}. Solving the eigenvalue problem by letting $\mathbf{F}=0$ in \cref{EQU C7: Rayleigh-Ritz} leads to the estimation of structural mode shapes and natural frequencies.

\subsubsection{Boundary condition identification by matching modal characteristics}
What modal information can be extracted from the reference impact data $Q, F(t), S(t)$? By analysing the structural response by mode superposition method, a key outcome is the identification of significant lower-mode natural frequencies. In addition to natural frequencies, transmissibility serves as another modal property that can be inferred from reference impact data. Two types of transmissibility are considered: sensor transmissibility and excitation transmissibility.

Sensor transmissibility is defined as the ratio of the spectral amplitudes of responses recorded by two sensors at a given natural frequency when the structure is excited by an impact. When piezoelectric (PZT) sensors monitor impacts and record strain responses, the sensor transmissibility between sensor $j$ and sensor $n$ across all reference impacts $i=1,...,N_i$  is given by: 
\begin{equation}
\begin{aligned} 
T^{jn}(\omega_m) 
= \frac{\sum_{i=1}^{N_i} \hat{s}(\omega_m;  q_i, L_j)}{\sum_{i=1}^{N_i} \hat{s}(\omega_m;  q_i, L_n)} 
= \frac{\sum_{i=1}^{N_i} \hat{h}(\omega_m; q_i, L_j)\hat{f}(\omega_m; q_i)}{\sum_{i=1}^{N_i} \hat{h}(\omega_m; q_i, L_n)\hat{f}(\omega_m; q_i)} 
= \frac{ \frac{\partial \phi^x_m}{\partial x}(L_j)+\frac{\partial \phi^y_m}{\partial y}(L_j) }{ \frac{\partial \phi^x_m}{\partial x}(L_n)+\frac{\partial \phi^y_m}{\partial y}(L_n) }, 
\end{aligned}
\end{equation}
where $\hat{s}(\omega; q_i, L_j)$ and $\hat{h}(\omega; q_i, L_j)$ denote the spectral response and transfer function at sensor location $L_j$ when excited at $q_i$, respectively. Here, $\phi^u_m$ denotes $m$-th mode shape in term of $u$-displacement, and $\frac{\partial \phi^u_m}{\partial x}(L_j)$ is its spatial derivative with respect to the $x$, evaluated at sensor location $L_j$. Essentially, sensor transmissibility reflects the ratio of mode shape gradients. However, sensor characteristics, including frequency response variations due to degradation or poor attachment, may introduce errors in the transmissibility estimate.

To address this limitation, excitation transmissibility is defined based on transfer functions and depends on the availability of reference impact data rather than sensor characteristics. Considering two independent impacts at locations $q_i$ and $q_k$, excitation transmissibility is defined as the ratio of the spectral transfer functions $\hat{h}(\omega; q_i, L)$ and $\hat{h}(\omega; q_k, L)$ at natural frequency $\omega_m$, evaluated across all sensors $L_j, j=1,...,N_s$:
\begin{equation}
\begin{aligned} \label{excitation transmissibility}
T_{ik}(\omega_m) 
= \frac{ \sum_{j=1}^{N_s} \hat{h}(\omega_m;  q_i, L_j)}{ \sum_{j=1}^{N_s} \hat{h}(\omega_m; q_k, L_j)} = \frac{ \phi_m(q_i)}{ \phi_m(q_k)}. 
\end{aligned}
\end{equation}
For both PZT sensors and accelerometers, excitation transmissibility directly corresponds to the ratio of mode shapes. This definition provides a more robust means of estimating modal properties, particularly in cases where sensor performance may be affected by degradation or attachment quality.

By leveraging both natural frequencies and transmissibility, boundary conditions can be identified by minimising the discrepancies between the extracted modal properties and those predicted using the Rayleigh-Ritz method. This optimisation problem is formulated as:
\begin{equation}
\begin{aligned} \label{EQU C7: boundary condition optimisation}
\Theta_{k} =  \underset{\Theta_k}{\arg\min} \; L_{nf} + \tau L_{et}, \;
L_{nf} = \sum_{m=1}^{N_{sm}}(\omega_m - \bar{\omega}_m)^2, \; L_{et} = \sum_{k=1}^{N_{i}} \sum_{i=k+1}^{N_{i}}  [T_{ik}(\omega_m) - \bar{T}_{ik}(\omega_m)]^2, 
\end{aligned}
\end{equation}
where $\Theta_{k}$ represents the stiffness of artificial springs applied at the boundaries. $L_{nf}$ and $L_{et}$ are the mean sqaure error loss of natural frequencies and excitation transmissibility, respectively. $N_{sm}$ denotes the number of significant identified modes from reference impact data. By optimising the artificial spring stiffness, the mode shapes and natural frequencies of the FSDT plate can be determined, facilitating the process of impact force identification.

\subsection{Physics-guided impact localisation and uncertainty quantification}
The GVPs $v_g(\theta; \omega)$ are identified by optimising the structural material properties in the FSDT modelling, allowing the generation of extended data and facilitating impact localisation. 

\subsubsection{Multi-fidelity Gaussian process regression for impact localisation}
The integration of physics-generated data enhances the generalisability of the model. By leveraging the identified GVPs $v_g(\theta; \omega)$, from the FSDT model, low-fidelity synthetic TDOA data $\Delta \mathbf{t}_{FSDT}(\omega_m)$ can be generated across the entire structure. Since these GVPs are inferred from experimental data, the simulated TDOAs maintain an intermediate level of accuracy, capturing the primary wave propagation characteristics while exhibiting some deviations from high-fidelity experimental data.

To improve the reliability of impact localisation, a multi-fidelity GPR framework \cite{forrester_multi-fidelity_2007} is employed, which integrates both simulated and experimental data. This approach enables accurate localisation across the structure by modelling the high-fidelity output as a scaled low-fidelity prediction with an additional correction term:
\begin{equation}
\begin{aligned} 
y_{HF}(\mathbf{x}) = \alpha y_{LF}(\mathbf{x}) + \delta(\mathbf{x})
\end{aligned}
\end{equation} 
where $y_{LF}(\mathbf{x})$ represents the low-fidelity model trained on the synthetic TDOA data $\Delta \mathbf{t}_{FSDT}(\omega_m)$. The scaling factor $\alpha$ adjusts for systematic discrepancies between low- and high-fidelity data, while the correction term $\delta(\mathbf{x})$ is modelled as a GP trained on sparse high-fidelity TDOA data from experimental measurements. This multi-fidelity learning framework effectively compensates for the limitations of the low-fidelity model, enabling robust and accurate impact localisation across the structure.

To effectively fuse multi-frequency TDOA data within a single GPR model, a tailored multi-frequency kernel is designed as a summation of uni-frequency radial basis function (RBF) kernel:
\begin{equation}
\begin{aligned} 
k(\mathbf{x}, \mathbf{x}') = \sum_{m=1}^{N_f}k_{rbf}(\mathbf{x}(\omega_m), \mathbf{x}'(\omega_m)). 
\end{aligned}
\end{equation} 
The RBF kernel $k_{rbf}$ contains a length-scale parameter: $l_{rbf}$. A key consideration is whether these length scales should be independently modelled for each frequency, particularly since TDOAs at higher frequencies tend to have lower magnitudes, significantly affecting the kernel values for a fixed $l_{rbf}$. However, increasing the number of hyperparameters in the GPR model elevates model complexity and raises the risk of overfitting.

The simulated impact locations are typically derived from structured experimental designs, such as uniform sampling over a 20 mm by 20 mm grid. Given the intermediate dataset size, reducing model complexity is crucial to achieving high generalisability. Therefore, it is preferable to model the multi-frequency composite kernel using a shared length scales $l_{rbf}$. To facilitate this, normalisation preprocessing of the TDOA data is required to ensure comparability across different frequencies. This normalisation is achieved using the dispersion relations $v_g(\omega; \theta)$:
\begin{equation}
\begin{aligned} 
\Delta \mathbf{t}_{nor}(\omega_m) ={\Delta \mathbf{t}(\omega_m)}{v_g(\omega_m; \theta=0)}.
\end{aligned}
\end{equation} 
By combining physics-augmented data generation with multi-frequency kernel design, the proposed GPR model is expected to achieve accurate and robust impact localisation with enhanced generalisability while also providing uncertainty quantification.

\subsection{Physics-guided impact force deconvolution and uncertainty quantification}
The force reconstruction integrates the constructed FSDT model and high-fidelity (HF) reference impact data. This integration enables estimation of the transfer function $\hat{h}(\omega; q, L)$ at a target location $q$, facilitating physics-guided impact force deconvolution via adaptive frequency-domain regularisation.

\subsubsection{Regularised force deconvolution in frequency domain}
Given the estimated transfer functions $\hat{h}(\omega; q, L)$ at the target location $q$, the impact force in the frequency domain can be reconstructed using the classical least-squares deconvolution:
\begin{equation}
\begin{aligned} 
\hat{f}(\omega; q) = \frac{ \sum_{j=1}^{N_s} \hat{s}(\omega; q, L_j) \hat{h}^*(\omega; q, L_j)}{\sum_{j=1}^{N_s} \left| \hat{h}(\omega; q, L_j) \right|^2},
\end{aligned}
\end{equation}
where $\left| \cdot \right|$ represents the absolute operator, $\hat{s}(\omega; q, L_j)$ is the j-th sensor response in the frequency domain, and $\hat{h}^*(\omega; q, L_j)$ is the complex conjugate of the transfer function.

However, this inversion problem is typically ill-posed that the possible presence of small values in $\hat{h}(\omega; q, L)$ can amplify noise and destabilise the solution. To mitigate this, regularisation techniques are employed to stabilise the inverse problem. A widely used stabilisation method is frequency-domain Tikhonov regularisation—also known as Wiener deconvolution \cite{martin_impact_1996, thiene_effects_2014, simone_hierarchical_2019}—which reformulates the inverse problem as a penalised least-squares optimisation:
\begin{equation}
\begin{aligned} 
\hat{f}(\omega; q, \lambda_{reg}) = \underset{\hat{f}(\omega; q)}{\mathrm{argmin}} \,  \sum_{j=1}^{N_s} \left\| \hat{s}(\omega; q, L_j) - \hat{f}(\omega; q)\hat{h}(\omega; q, L_j) \right\|^2 + \lambda_{reg}\left\| \hat{f}(\omega; q) \right\|^2, 
\end{aligned}
\end{equation}
where $\left\| \cdot \right\|$ represent the Euclidean norm operator, $\lambda_{reg}$ is the regularisation parameter that imposes a global smoothing effect. This penalisation yields a closed-form solution: 
\begin{equation}
\begin{aligned} \label{EQU: reg deconvolution frequency domain}
\hat{f}(\omega; q, \lambda_{reg}) = \frac{\sum_{j=1}^{N_s} \hat{s}(\omega; q, L_j) \hat{h}^*(\omega; q, L_j)}{\sum_{j=1}^{N_s} \left| \hat{h}(\omega; q, L_j) \right|^2+\lambda_{reg}}.
\end{aligned}
\end{equation}

The regularisation parameter $\lambda_{reg}$ controls the trade-off between fidelity to the measured signal and smoothness of the reconstructed force. Its selection is critical and can be automated using data-driven approaches such as the L-curve criterion \cite{qiu_impact_2019, chen_comparative_2021} or Generalised Cross Validation (GCV) \cite{wang_bandlimited_2020, choi_comparison_2007}. GCV, in particular, provides a principled and automated approach for selecting $\lambda_{reg}$ without requiring ground truth information. The GCV functional for frequency-domain deconvolution across multiple sensors is given by:
\begin{equation}
\begin{aligned} \label{EQU GCV}
\mathrm{GCV}(\lambda_{reg}) = \frac{\sum_{j=1}^{N_s} \sum_{k=1}^{N_f}\left| \frac{\lambda_{reg}}{\left| \hat{h}(\omega_k;q,L_j)\right|^2 + \lambda_{reg}} \hat{s}(\omega_k;q,L_j) \right|^2  }{ \left( \sum_{j=1}^{N_s} \sum_{k=1}^{N_f} \frac{\lambda_{reg}}{\left| \hat{h}(\omega_k;q,L_j)\right|^2 + \lambda_{reg}} \right)^2 }, 
\end{aligned}
\end{equation}
with the optimal $\lambda_{reg}$ obtained by minimising the above expression.

While $\ell_2$ regularisation is effective in stabilising the inversion, it often leads to underestimation of the true force magnitude \cite{qiao_non-convex_2020}, especially at low frequencies where penalisation is overly aggressive. To mitigate this, frequency-dependent regularisation schemes are adopted, which allow adaptive penalisation across the spectrum and improve reconstruction fidelity.

In addition to frequency-domain $\ell_2$-based methods, several time-domain techniques exploit the temporal sparsity of impact forces, such as sparse regularisation ($\ell_1$-based) \cite{qiao_sparse_2017, liu_impact_2020} and wavelet-based regularisation \cite{tran_development_2018, xiao_impact_2024-1}. These methods offer better temporal resolution and robustness, especially when only a few sensor signals are available. Nonetheless, both $\ell_1$ and $\ell_2$ regularisers tend to underestimate force magnitudes \cite{qiao_non-convex_2020}, and their accuracy is constrained when frequency response variability is significant.

\subsubsection{Fidelity of transfer function estimation from FSDT and reference impact data}
The data-driven FSDT model is calibrated via sequential estimation of material properties and boundary conditions. However, it cannot perfectly replicate sensor responses to actual impacts due to several limitations:
\begin{itemize}
\item Model simplifications: The FSDT plate model neglects transverse normal strain and assumes a constant transverse shear strain through the plate thickness. Additionally, the boundary conditions are modelled using uniform spring stiffness along each edge. These simplifications limit its ability to fully capture the real-world structural dynamics.
\item Sparse reference data: The accuracy of material property and boundary condition identification depends on the availability of reference data, which may be insufficiently dense to ensure precise reconstruction.
\item Sensor frequency responses function variability: The sensor’s inherent frequency responses function is affected by factors such as degradation and attachment quality, introducing further discrepancies.
\end{itemize}

Therefore, while the FSDT model provides a low-fidelity (LF) approximation of the transfer function, reference impacts offer high-fidelity (HF) information. Impact force reconstruction must prioritise HF data, with LF models used to provide physical constraints or regularisation guidance.

\subsubsection{Strategies for multi-fidelity transfer function estimation}
To estimate the transfer function $\hat{h}(\omega; q, L)$ at a new location $q$, the following strategies combine LF physics and HF reference data:
\begin{enumerate}
\item Physical prior approach: Utilise the transfer function estimated by the FSDT model as an LF physical prior, while the HF reference impact data are employed to model the residual between the physical prior and the reference measurements.
\item Modal kernel design: Incorporate physical modal characteristics into GPR or RBF interpolation \cite{simone_hierarchical_2019}. For instance, integrating modal similarity into the GPR kernel ensures that interpolation is influenced not only by spatial proximity but also by modal resemblance, thereby improving accuracy.
\item RBF interpolation \cite{simone_hierarchical_2019} with adaptive regularisation: Perform interpolation/extrapolation of the transfer function separately using both HF reference transfer functions and LF FSDT transfer functions. An adaptive regularisation factor is introduced, defined based on the discrepancy between the interpolated FSDT results and theoretical FSDT predictions.
\end{enumerate}

The choice among these depends on the FSDT model’s fidelity and reference impact coverage:
\begin{itemize}
\item Strategy 1 is preferable for accurate FSDT models with sparse reference data. 
\item Strategy 2 improves interpolation accuracy but requires careful kernel design and a reliable modal basis. 
\item Strategy 3 offers a flexible compromise when FSDT accuracy is moderate and HF data are sparse.
\end{itemize}
In this work, Strategy 3 is adopted due to the intermediate-to-low accuracy of the data-driven FSDT model. While the LF model cannot be used directly as a prior mean or to calibrate HF estimates, it can still provide valuable information about the spatial complexity of local dynamics and the frequency-dependent quality of interpolation. Therefore, it is utilised to guide adaptive regularisation in force deconvolution.

\subsubsection{Transfer function estimation and force deconvolution with adaptive regularisation}
Let the HF and LF transfer function estimates at the reference impact locations $Q$ be denoted by:
\begin{itemize}
    \item $\hat{H}(\omega; Q, L)$: from experimental data;
    \item $\hat{H}_{FSDT}(\omega; Q, L)$: from the FSDT model.
\end{itemize}
Using RBF interpolation, the estimated transfer functions at a new location $q$ are denoted as:
\begin{itemize}
    \item $\hat{h}^{RBF}(\omega; q, L)$: interpolated from HF reference data,
    \item $\hat{h}_{FSDT}^{RBF}(\omega; q, L)$: interpolated from LF FSDT data.
\end{itemize}
The explicit RBF formulations for these interpolants are provided in \cref{subapp: RBF}, following the methodology in \cite{simone_hierarchical_2019}. Briefly, each interpolant combines a weighted sum of radial basis functions and a low-order polynomial to ensure well-posedness and smoothness:
\begin{equation}
\begin{aligned}
&\hat{h}^{RBF}(\omega; q, L) = \sum_{i=1}^{N_i} \lambda_i\psi(\left|q - Q_i \right|) + \sum_{j=1}^{M} \varsigma_j p_j(q), \\
&\hat{h}^{RBF}_{FSDT}(\omega; q, L) = \sum_{i=1}^{N_i} \lambda_i^{FSDT}\psi(\left|q - Q_i \right|) + \sum_{j=1}^{M} \varsigma_j^{FSDT} p_j(q). 
\end{aligned}
\end{equation}
where $\psi(\cdot)$ represents the chosen radial basis function, and $p_j(\cdot)$ denotes a polynomial term included to ensure the well-posedness of the RBF interpolation.

The theoretical transfer function predicted by the FSDT model at location $q$ is denoted as $\hat{h}_{FSDT}(\omega; q, L)$. The discrepancy between the RBF-interpolated transfer function and the theoretical FSDT prediction is quantified by their difference:
\begin{equation} 
\begin{aligned} \label{EQU C7: adaptive R}
\lambda_{reg}(\omega) = \cumsum \left(  \frac{1}{N_s}\sum_{j=1}^{N_s}\left|\hat{h}^{RBF}_{FSDT}(\omega; q, L_j)  - {\hat{h}_{FSDT}(\omega; q, L_j)} \right|^2 \right),
\end{aligned}
\end{equation}
where $\cumsum$ denotes the cumulative sum operator applied across the frequency range. This adaptive, frequency-dependent regularisation term, $\lambda_{reg}(\omega)$, reflects the degree of mismatch between the physics-based (FSDT) model and the data-driven RBF interpolation. By incorporating this discrepancy into the regularisation term of the frequency-domain deconvolution formulation (see \cref{EQU: reg deconvolution frequency domain}), the solution is penalised more heavily in frequency bands where the interpolation error is high, thereby improving the robustness and accuracy of impact force reconstruction.

\section{Experimental validation by impact testing}  \label{section exprimental validation}
\begin{figure}[htb] 
	\centering
		\includegraphics[width=0.9\columnwidth]{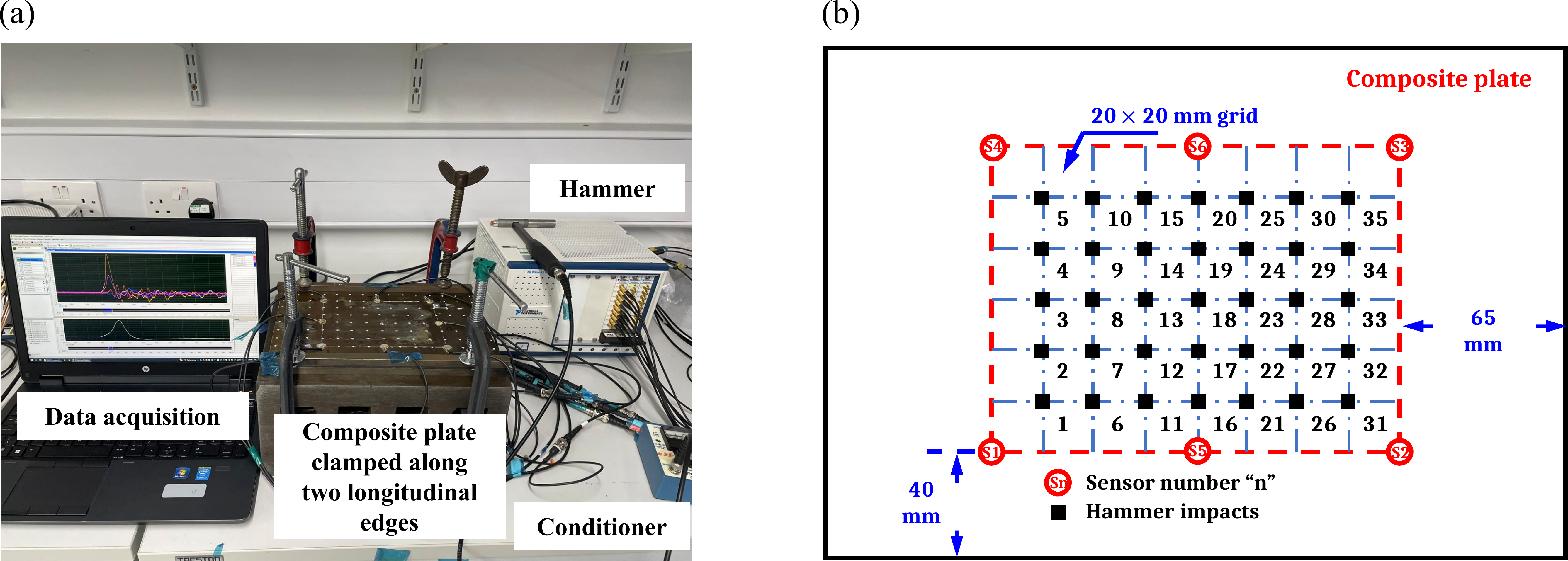}
	\caption{Impact testing using handheld hammer.}
	\label{FIG C7: exp_testing}
\end{figure}

Experimental impact tests were conducted to capture the structural response and impact force history of a sensorised fibre-reinforced composite plate subjected to hammer impacts, as illustrated in \cref{FIG C7: exp_testing}(a).
The composite plate, fabricated from M21/T800 prepreg material, measured $290 \times 200 \times 4 $ mm and had a quasi-isotropic layup of $[0/+45/-45/90]_{2s}$. During impact testing, this composite plate was clamped along its two longer edges, leaving the two shorter edges free. Impact excitation was generated using a PCB Piezotronics 086C03 hammer, which has a weight of 160 g. The layout of the composite plate and the hammer impact locations are illustrated in \cref{FIG C7: exp_testing}(b), where the impact points are marked as black solid squares. These impacts were applied at the vertices of a 20 $\times$ 20 mm grid. During each hammer strike, four PZT sensors recorded the structural responses, while the ICP quartz force sensor integrated within the hammer measured the impact force history. For reference, detailed reference elastic properties of M21/T800 lamina are listed in \cite{xiao_hybrid_2024, xiao_general_2025}. As this study focuses on a purely data-driven approach, these reference material properties are used solely for the accuracy validation of the estimated properties.

\section{Impact identification results and analysis}  \label{section Results and Analysis}
\subsection{Sparse data-driven FSDT plate modelling} 
\subsubsection{Material property identification}
As demonstrated in the authors' previous study \cite{xiao_general_2025}, the combination of four reference impacts and a sparse sensor network consisting of four PZT sensors enables the identification of GVPs with high accuracy. Given this effectiveness, the same configuration is expected to yield reliable estimates of the material properties. Considering the dominance of low-frequency modes in impact dynamics, material property identification is focused on the dispersion characteristics within the frequency range of 1–10 kHz. By matching the estimated dispersion relations with experimental measurements obtained from the four reference impacts, the material properties of the FSDT plate are identified through the maximisation of the log-likelihood function defined in \cite{xiao_general_2025}.

\begin{figure}[htb] 
	\centering
		\includegraphics[width=1\columnwidth]{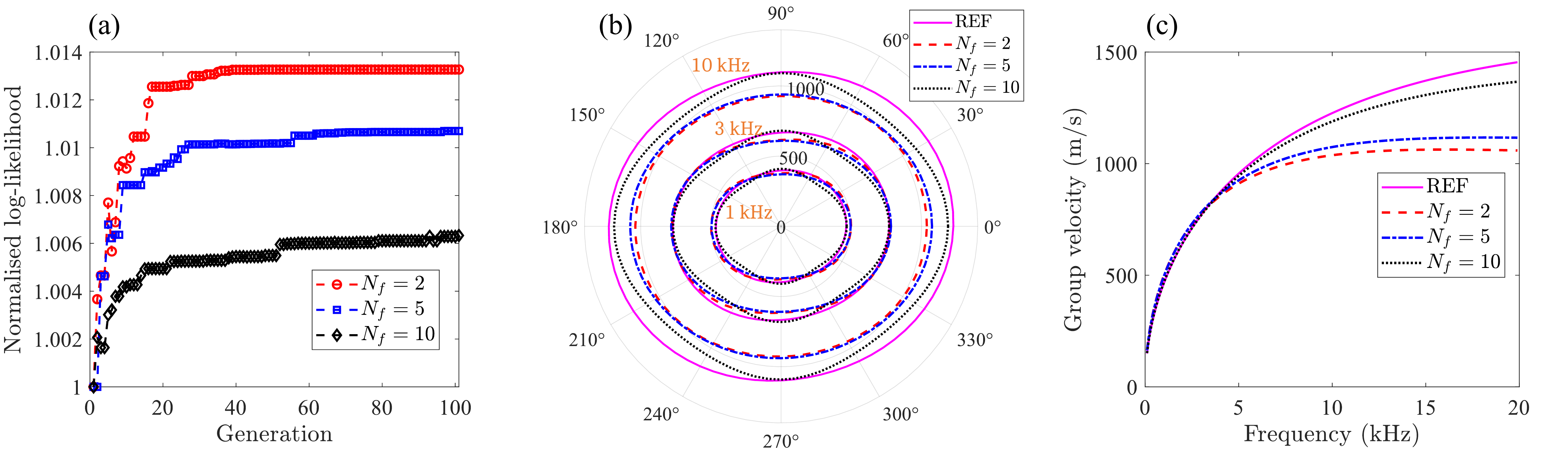}
	\caption{Material properties identification with four reference impacts. }
	\label{FIG: material properties identification}
\end{figure}

As illustrated in \cref{FIG: material properties identification}(a), the log-likelihood function is maximised based on four reference impacts numbered 12, 14, 22, 24 shown in \cref{FIG C7: exp_testing}(b), across three frequency ranges, where $N_f$ denotes the number of frequency points considered, spanning from 1 kHz to $N_f$ kHz in 1 kHz increments. With 100 generations in the GA, all three cases ($N_f=2$, $N_f=5$, and $N_f=10$) converge to a stable maximum.

The comparison between the GVPs derived from the optimised material properties and those corresponding to the reference properties is presented in \cref{FIG: material properties identification}(b), where the reference GVPs are depicted as solid magenta lines and the optimised GVPs as dashed lines. Among the three cases, the $N_f=10$ optimisation yields the most accurate GVPs, as the inclusion of a wider frequency range allows the estimated GVPs to align more closely with those obtained from the reference material properties.

A noticeable trend is observed: as the frequency increases, the reference GVPs tend to be higher than the optimised ones. This discrepancy can be attributed to two main factors: (1) the FSDT model inherently overestimates GVPs at higher frequencies compared to experimental measurements, and (2) at lower frequencies, where the TDOA values are larger, the optimisation process prioritises the low-frequency components to maximise the likelihood function. The comparison of dispersion relations at $\theta=0^\circ$ between the optimised and reference models is illustrated in \cref{FIG: material properties identification}(c). For $N_f=2$ and $N_f=5$, significant deviations from the reference dispersion relations occur at higher frequencies, whereas for $N_f=10$, the estimated dispersion relations remain more consistent across the entire frequency range. This improvement can be attributed to the integration of a wider frequency spectrum in the optimisation, leading to enhanced accuracy in the high-frequency range.

Despite the observed discrepancies between the reference and optimised GVPs, particularly at higher frequencies, the optimised GVPs remain highly accurate for impact localisation, especially in the low-frequency range. This is because they are directly optimised based on reference experimental impact data, ensuring reliability in practical applications.

\cref{TB C7: optimised plate-level material properties} summarises the identified plate-level material properties for different frequency ranges. The reference values (REF) represent the known material properties of the composite plate, while the optimised values correspond to different $N_f$ cases. The density ($\rho$) and plate thickness ($d$) remain relatively consistent across the cases, exhibiting only slight variations due to the optimisation process. However, the bending stiffness components ($D_{11}$, $D_{22}$, and $D_{66}$) show more pronounced differences, particularly for $D_{11}$, which exceeds the reference value.

These variations highlight the inherent complexity of the optimisation problem, which features multiple local minima that yield dispersion relations and GVPs closely matching experimental data. Given that experimental measurements may not perfectly align with the theoretical reference properties due to factors such as manufacturing variations and measurement noise, achieving an exact match with the reference material properties is nearly impossible. Instead, the optimisation process converges to a local minimum that ensures highly accurate GVPs, particularly in the low-frequency range, where wave propagation characteristics are most critical for impact localisation.

\begin{table}[ht]
\centering
\captionsetup{justification=centering}
\caption{Identified plate-level material properties. } 
\label{TB C7: optimised plate-level material properties}
{\small 
\renewcommand{\arraystretch}{1.45} 
\begin{tabular}{lllllll}
\toprule
Cases &
$\rho$ ($\mathrm{kg/m^3}$) &
$d$ (mm) &
$I_{1}$ ($\mathrm{kg/m^2}$) &
$D_{11}$ ($\mathrm{Pa\cdot m^3}$)&
$D_{22}$ ($\mathrm{Pa\cdot m^3}$)&
$D_{66}$ ($\mathrm{Pa\cdot m^3}$)\\

REF &
1600 &
4.00 &
6.40 &
511.0 &
269.9 &
125.4 \\

$N_f=2$ &
1700 &
3.92 &
6.67 &
755.1 &
198.6 &
179.5 \\

$N_f=5$ &
1543.8 &
4.20 &
6.48 &
692.8 &
191.8 &
114.5 \\

$N_f=10$ &
1602.2 &
4.19&
6.73 &
820.3 &
396.3 &
21.5\\
\bottomrule
\end{tabular} \\
}
\end{table}

\subsubsection{Boundary Condition Identification}
The modal characteristics of a structure can be estimated from the responses of PZT sensors and transfer functions when the impact forces are known. Impact forces, such as half-sine impulses, exhibit their highest spectral amplitude in the low-frequency range. This amplifies the contribution of low-frequency components in the PZT sensor signals, making it easier to identify low-frequency modes from these signals compared to transfer functions. In contrast, transfer functions tend to exhibit an energy shift towards higher frequencies, facilitating the identification of higher-frequency modes.

\begin{figure}[htb] 
	\centering
		\includegraphics[width=0.8\columnwidth]{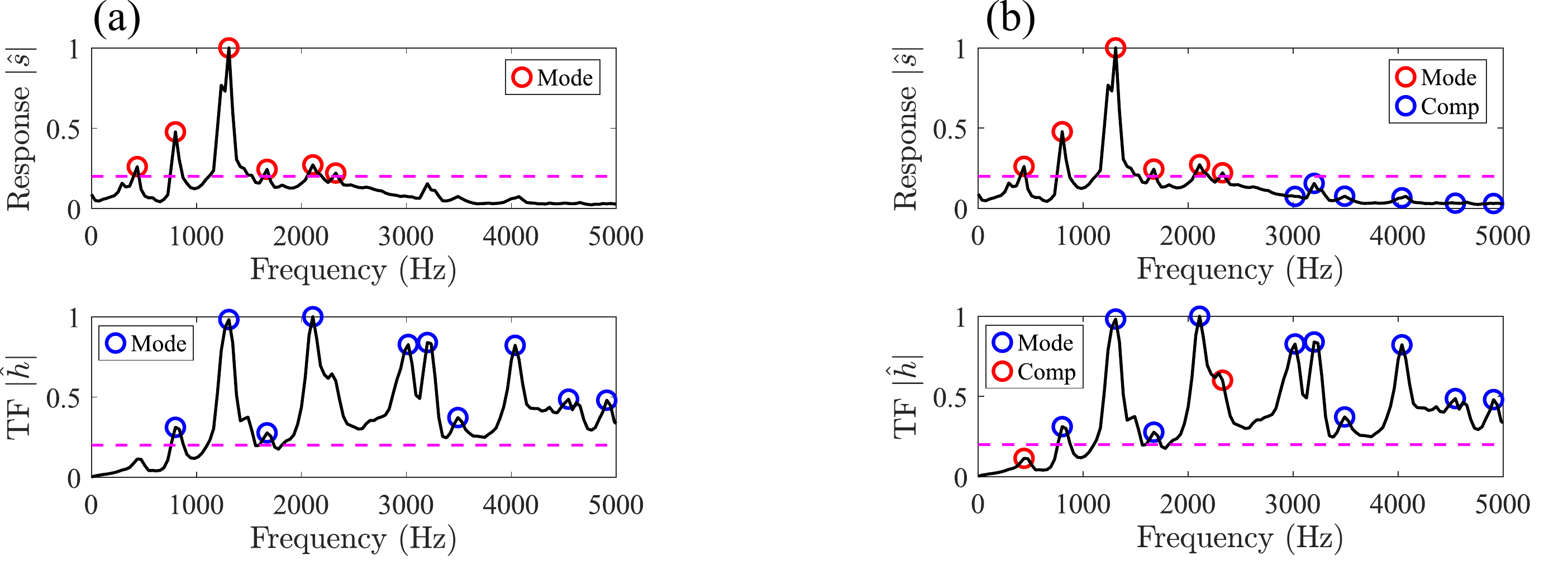}
	\caption{Mode identification from PZT sensor signals and transfer functions. }
	\label{FIG C7: sig_tf_mode}
\end{figure}

To ensure that each reference impact contributes equally to the modal identification process, the spectra and transfer functions of the reference impacts are normalised before averaging. \cref{FIG C7: sig_tf_mode}(a) illustrates mode identification using PZT sensor signals and transfer functions obtained from the sparse reference impact data. A comparison of the spectral distributions of both the PZT sensor response and the transfer function (TF) reveals that lower-frequency components (below 2000 Hz) dominate the PZT sensor responses. The relatively low spectral amplitude at higher frequencies makes it challenging to accurately identify high-frequency modes from PZT sensor signals alone. In contrast, the transfer function exhibits a spectral energy shift towards higher frequencies, enabling the identification of high-frequency modes while struggling to capture the lowest mode due to a small peak.

The PZT sensor response and the transfer function provide complementary modal information. As shown in \cref{FIG C7: sig_tf_mode}(b), combining both methods allows for the identification of modes across a broader frequency range, capturing both low-frequency and high-frequency modes more effectively. This combined approach enhances the accuracy of mode identification, which is crucial for excitation transmissibility estimation and subsequent boundary condition identification.

As shown in \cref{FIG C7: exp_testing}, the composite plate was clamped along its longer edges (though not perfectly) while the shorter edges remained free. To model these boundary conditions in a simplified manner, artificial translational and rotational springs were introduced at the plate edges. The design variables consist of four spring stiffness values, one for each edge, with the assumption that the translational and rotational stiffnesses are identical for a given edge.

Although the two shorter edges of the experimental plate are nominally free, their stiffness values were still included in the model, constrained within the range [0, 0.5]. Meanwhile, the stiffness values of the two clamped edges were bounded within [0, 1]. To ensure numerical stability and avoid excessively large stiffness values in optimisation, the stiffness parameters were normalised using the transformation:
\begin{equation} 
\begin{aligned}
k_{nor} = \frac{\log k}{10},
\end{aligned}
\end{equation}
where a maximum real stiffness value of 1e10 corresponds to a perfectly clamped edge. Beyond this value, further increases in stiffness do not affect the simulated natural frequencies in the FSDT model. 

\begin{figure}[htb] 
	\centering
		\includegraphics[width=0.45\columnwidth]{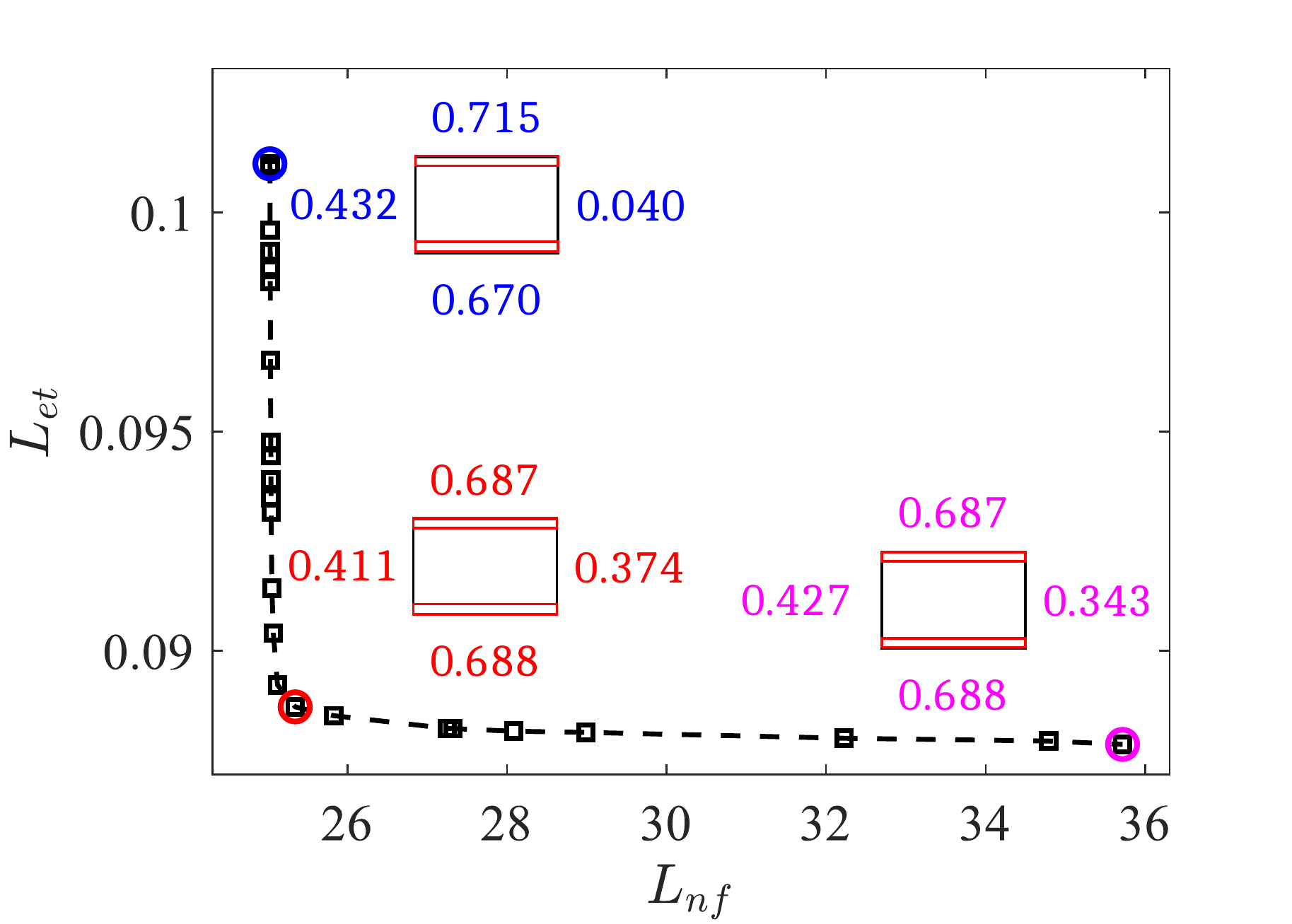}
	\caption{Boundary condition identification. }
	\label{FIG C7: BC identification}
\end{figure}

\cref{FIG C7: BC identification} presents the Pareto front of the minimisation process, balancing the natural frequency loss $L_{nf}$ and excitation transmissibility loss $L_{et}$. The Pareto front exhibits an L-curve, where significant variations in one loss function occur along the two distinct edges of the 'L', while the other loss remains nearly unchanged. This characteristic suggests that the optimal solution should be selected at the transition point of the L-curve, where a balanced trade-off between the two loss functions is achieved.

At the transition point, marked by the red circle in \cref{FIG C7: BC identification}, the identified normalised stiffness values for the two longitudinal edges (clamped but not perfectly rigid) were 0.687 and 0.688. These correspond to real stiffness values more than 100 times greater than those of the two shorter edges, which were nominally free. This indicates that the longitudinal edges serve as the effective boundary conditions, while the shorter edges have a negligible influence. The identified boundary condition configuration aligns well with the experimental setup, thereby validating the effectiveness of the proposed boundary condition identification method.

To further validate the identified boundary conditions, \cref{TB C7: identified natural frequencies} presents the natural frequencies of the first 10 modes, extracted from the reference impact data and compared with those identified using the optimised FSDT model. The results indicate an average discrepancy of approximately 80 Hz between the experimentally extracted and model-predicted natural frequencies. This frequency deviation suggests that while the identified boundary conditions provide a reasonable approximation, the transfer function estimated from the constructed FSDT model may exhibit slight peak shifts compared to the experimentally obtained transfer function from reference impact data. These discrepancies may arise from unmodelled factors such as localised boundary flexibility, material inhomogeneities, or additional damping effects, which could be further refined in future studies.
\begin{table}[ht]
\centering
\captionsetup{justification=centering}
\caption{Identified natural frequencies.} 
\label{TB C7: identified natural frequencies}
{\small 
\renewcommand{\arraystretch}{1.45} 
\begin{tabular}{lllllllllllll}
\toprule
Mode &
1 &
2 &
3 &
4 &
5 &
6 &
7 &
8 &
9 &
10 \\

Extracted $\omega_m$  (Hz) &
436.4 &
800.0 &
1309.1 &
1672.7 &
2109.1 &
2327.3 &
2727.3 &
3018.2 &
3200.0 &
3490.9 \\

Estimated $\bar{\omega}_m$ (Hz) &
459.9 &
726.0 &
1272.0 &
1545.0 &
1887.2 &
2371.7 &
2774.5 &
3032.4 &
3273.8 &
3514.6 \\
\bottomrule
\end{tabular} \\
}
\end{table}

\subsection{Impact localisation based on physics-augmented multi-fidelity GPR} 
\label{subsection Impact localisation}
Based on sparse experimental reference impact data and the identified GVPs (approximated physics), a GPR model was constructed and trained to locate a total of 35 experimental impacts on the composite plate. The approximated physics enables the generation of low-fidelity (LF) physics-augmented data, which provides additional training information for the GPR model. For the target structure, these physics-augmented data points were generated using uniform sampling on a 20 $\times$ 20 mm grid across the structure, ensuring comprehensive spatial coverage.

\subsubsection{Impact localisation: low-fidelity, high-fidelity and multi-fidelity}
\begin{figure}[htb] 
	\centering
		\includegraphics[width=1\columnwidth]{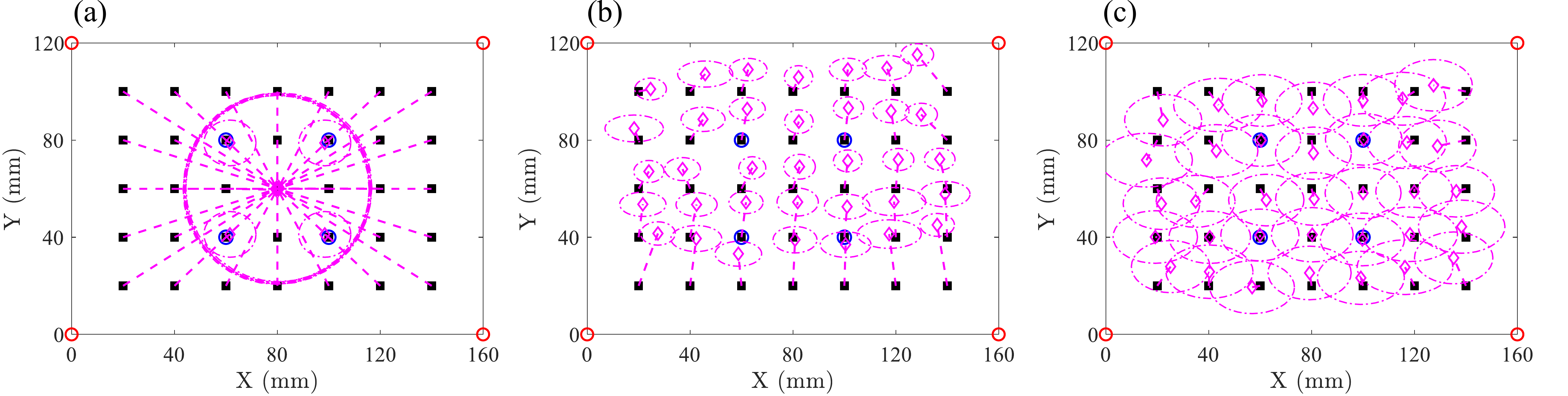}
	\caption{Impact localisation based on three approaches: (a) high-fidelity estimation based on sparse experimental impact data, (b) low-fidelity estimation based on physics-augmented data, (b) multi-fidelity estimation combining sparse experimental and physics-augmented data.}
	\label{FIG C7: loc_fidelity}
\end{figure}

To evaluate the influence of the identified GVPs (approximated physics) on impact localisation performance, three different localisation approaches were compared: (1) high-fidelity (HF) localisation using only sparse experimental impact data, (2) low-fidelity (LF) localisation using solely physics-augmented data, and (3) multi-fidelity (MF) localisation incorporating both experimental and physics-augmented data. \cref{FIG C7: loc_fidelity} illustrates the localisation results for these three approaches, specifically for the frequency range $N_f = 2 \to 2$. Here, the notation $N_f = {f_1} \to {f_2}$ indicates that the GVPs (material properties) were identified based on dispersion relations within the frequency range of 1 kHz to ${f_1}$ kHz, while the impact localisation process utilised GVPs derived from the frequency range of 1 kHz to ${f_2}$ kHz. 

The HF GPR model, trained exclusively on four sparse experimental reference impacts, struggles to capture the underlying wave propagation physics that govern the relationship between input TDOA and impact location. Due to data sparsity, the model adopts a constant mean function based on the reference impact data, with a large characteristic length scale. Consequently, the covariance function contributes minimally compared to the noise variance, causing the RBF kernel to fail in capturing variations in the data. Consequently, as depicted in \cref{FIG C7: loc_fidelity}(a), while the HF GPR model exhibits high accuracy for localising the reference impacts, it lacks extrapolation capability and defaults to predicting impacts outside the reference impact region at the mean of the reference impacts.

By contrast, the LF GPR model, trained on the densely sampled physics-augmented data, achieves intermediate accuracy but significantly improved generalisability across the structure, as shown in \cref{FIG C7: loc_fidelity}(b). Since these physics-augmented data points are generated using the approximated GVPs, the LF GPR model can predict impact locations across the entire structure. However, due to the imperfection of the approximated physics and the absence of experimental reference data, the LF GPR model maintains only moderate accuracy, even for the reference impacts used to approximate the GVP physics. 

The MF GPR model integrates the strengths of both approaches, combining the high generalisability of physics-augmented data with the high accuracy of sparse experimental reference impacts. As shown in \cref{FIG C7: loc_fidelity}(c), this hybrid approach achieves both high localisation accuracy and strong extrapolation capability, demonstrating superior performance across the entire structure. The combination of experimental and physics-augmented data allows the model to refine its predictions, ensuring both precision near reference impacts and reliable localisation in unexplored regions. 

\subsubsection{Impact localisation: variability in frequency ranges}
\begin{figure}[htb] 
	\centering
		\includegraphics[width=1\columnwidth]{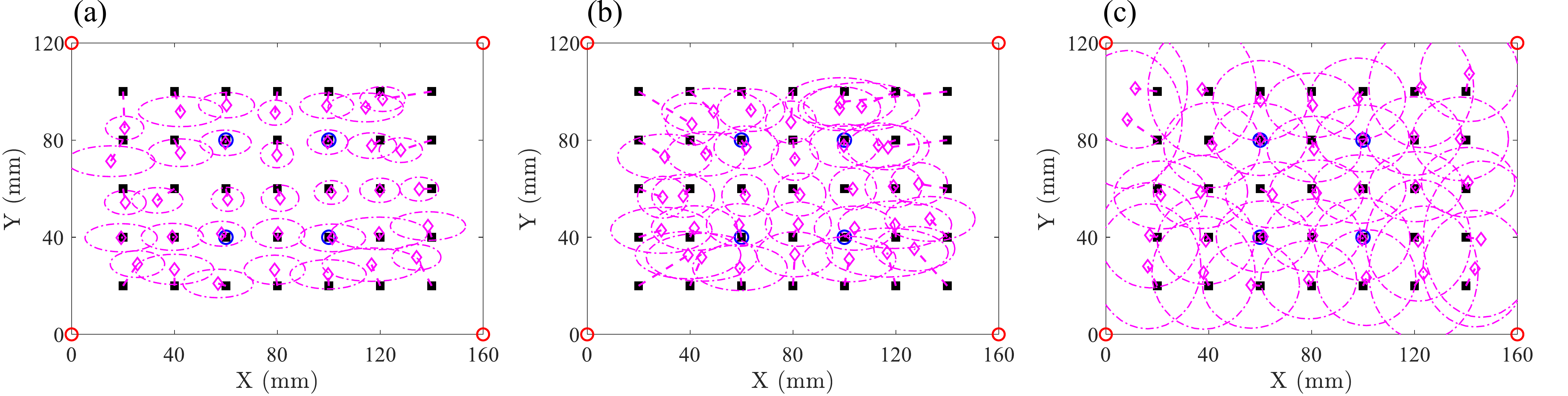}
	\caption{Multi-fidelity impact localisation based on different frequency ranges: (a) $N_f = 5 \to 2$, (b) $N_f = 10 \to 2$, (c) $N_f = 5 \to 5$. }
	\label{FIG C7: loc_vary_frequency}
\end{figure}

The notation $N_f = {f_1} \to {f_2}$ underscores the impact of frequency range selection on both material property identification and impact localisation accuracy. \cref{FIG C7: loc_vary_frequency} presents the results of multi-fidelity impact localisation using different frequency ranges. A comparison between localisation results for $N_f = 5 \to 2$ in \cref{FIG C7: loc_vary_frequency}(a), $N_f = 10 \to 2$ in \cref{FIG C7: loc_vary_frequency}(b) and $N_f = 2 \to 2$ in \cref{FIG C7: loc_fidelity}(c) reveals that the $N_f = 2 \to 2$ case achieves the highest localisation accuracy. In this case, nearly all predicted impact locations fall within the 95 $\%$ localisation confidence region, denoted by the dashed ellipse. This improvement can be attributed to the fact that when ${f_1}={f_2}=2$, the identified GVPs at these frequencies align most closely with the experimental TDOA data, leading to the most accurate impact localisation. Conversely, when ${f_1}> {f_2}=2$, higher-frequency components contribute to GVP identification but are not utilised in impact localisation, which reduces the accuracy of the estimated GVPs at low frequencies and consequently degrades impact localisation performance.

Additionally, comparing $N_f = 2 \to 2$ in \cref{FIG C7: loc_fidelity}(c) to $N_f = 5 \to 5$ in \cref{FIG C7: loc_vary_frequency}(c) further demonstrates that a lower frequency range yields more robust and reliable impact localisation. This observation stems from the fact that higher-frequency GVPs correspond to higher wave velocities, resulting in smaller TDOA values that are more sensitive to random noise. The increased influence of noise at higher frequencies propagates through the GVP identification process, leading to greater uncertainties in the estimated GVPs and, subsequently, in impact localisation. Therefore, it is recommended to prioritise lower-frequency ranges for impact localisation to enhance stability and accuracy.

\subsubsection{Impact localisation: variability in reference impacts}
By carefully selecting the reference impact locations, impact identification can be reformulated as a pure interpolation problem. In contrast, in the previous setup, most impact locations (numbers 1–35) fell outside the four reference impact points, introducing extrapolation errors. As demonstrated in \cite{xiao_robust_2025}, using nine strategically chosen reference impacts—numbered [1,3,5,16,18,20,31,33,35]—enables accurate interpolation-based impact localisation, achieving a mean localisation error of approximately 10 mm. 

\begin{figure}[htb] 
	\centering
		\includegraphics[width=1\columnwidth]{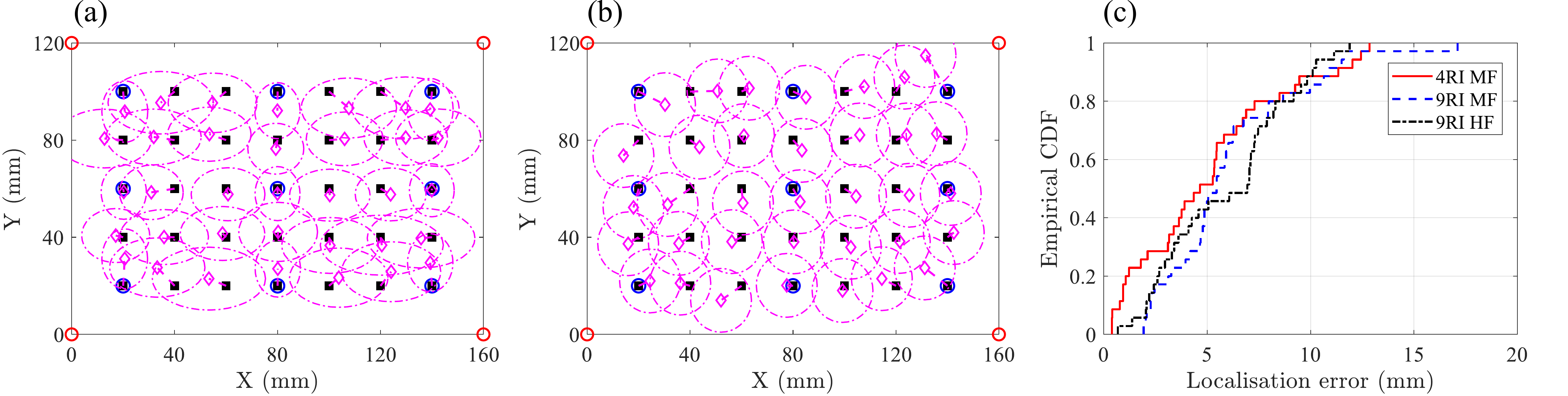}
	\caption{Impact localisation with nine reference impacts and $N_f = 5 \to 2$: (a) high-fidelity estimation, (b) multi-fidelity estimation, (c) empirical CDF of localisation error.}
	\label{FIG C7: loc_vary_RI}
\end{figure}

These nine HF reference impacts were utilised for material property identification, facilitating physics-augmented MF impact localisation. \cref{FIG C7: loc_vary_RI} presents the localisation results with these nine reference impacts and a frequency reduction from $N_f = 5 \to 2$. As shown in \cref{FIG C7: loc_vary_RI}(a), HF estimation based solely on the nine HF experimental reference impacts achieves a high localisation accuracy, with a mean error of 5.9 mm. In comparison, the MF estimation, which integrates the identified GVP physics with the same nine HF reference impacts, results in a mean localisation error of 6.0 mm. As expected, increasing the number of HF data points enhances interpolation accuracy within the reference coverage area. However, the MF model, by incorporating physics-based constraints, improves extrapolation performance beyond the reference coverage.

Compared to MF estimation using only four HF reference impacts, as illustrated in \cref{FIG C7: loc_fidelity}(c) and \cref{FIG C7: loc_vary_RI}(c), the MF model with four HF reference impacts achieves a mean localisation error of 5.0 mm, which is comparable to the HF and MF estimations using nine HF reference impacts. This result highlights the efficiency of the proposed MF approach, demonstrating that accurate localisation can be achieved with a minimal amount of HF data, thereby reducing experimental costs while maintaining high performance.

\subsection{Impact force reconstruction based on physics-augmented RBF}
\label{subsection Impact force reconstruction}
\subsubsection{Low-fidelity transfer functions from FSDT model}
\begin{figure}[htb] 
	\centering
		\includegraphics[width=0.8\columnwidth]{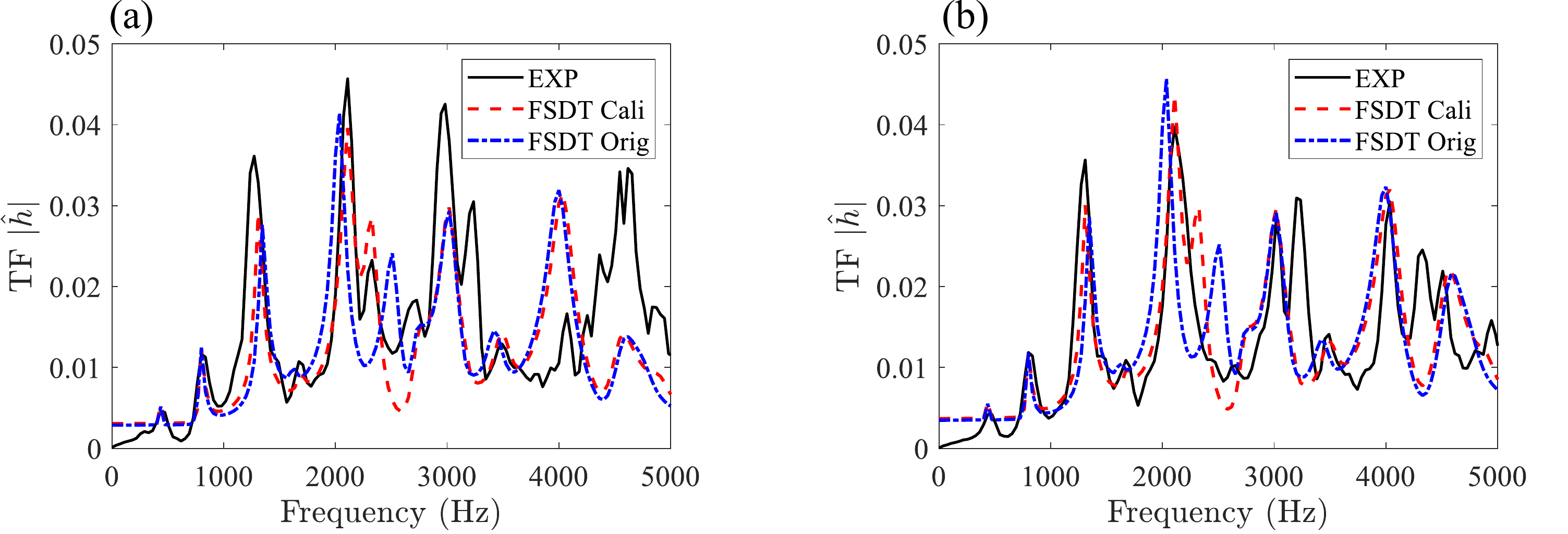}
	\caption{Transfer function estimates based on FSDT model: (a) reference impact 12, (a) reference impact 14. }
	\label{FIG C7: TF_identified}
\end{figure}

With the completion of the FSDT model through material property and boundary condition identification, the system transfer function can be derived from the modal characteristics using the principle of superposition. \cref{FIG C7: TF_identified} presents the mean transfer function estimates across four sensors for two reference impacts, numbered 12 in (a) and 14 in (b), based on the constructed FSDT model. However, as indicated in \cref{TB C7: identified natural frequencies}, discrepancies exist between the identified natural frequencies and the actual values extracted from the reference impact data. Consequently, the transfer function estimated from the original FSDT model (denoted as 'FSDT Orig' in the figure legend) exhibits slight peak shifts when compared to the experimentally obtained transfer function from reference impact data (denoted as 'EXP').

To improve the accuracy of the transfer function, the FSDT model was further calibrated by adjusting the identified natural frequencies to match their experimentally obtained counterparts. This calibration process ensures consistency in the frequency domain, aligning the peak locations of the modelled transfer function with those of the experimental data. As shown in \cref{FIG C7: TF_identified}, the calibrated FSDT transfer function ('FSDT Cali') demonstrates a significantly improved match with the experimental data in terms of peak positions.

However, despite this calibration, notable discrepancies persist in the peak magnitudes, particularly at higher frequencies. These deviations can be attributed to the simplified assumptions in the FSDT model and the errors in the data-driven FSDT modelling. Consequently, even after calibration, the transfer function estimates derived from the FSDT model remain low-fidelity approximations, in contrast to the high-fidelity experimental reference data. 

\subsubsection{Transfer function interpolation/extrapolation using RBF method}
\begin{figure}[htb] 
	\centering
		\includegraphics[width=0.8\columnwidth]{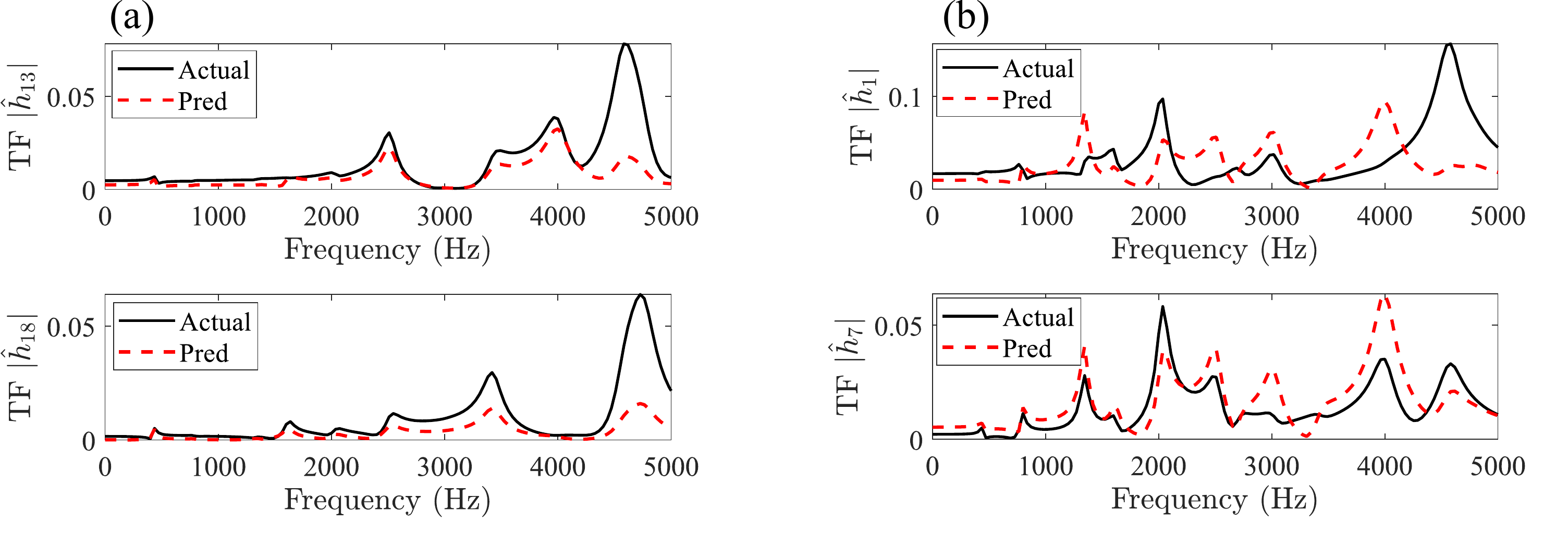}
	\caption{Transfer function interpolation/extrapolation based on FSDT model: (a) interpolation of impact 13 and 18, (a) extrapolation of impact 1 and 7. }
	\label{FIG C7: TF_FSDT_RBF}
\end{figure}

In location-based transfer function interpolation and extrapolation, the accuracy of the RBF method is highly dependent on the spatial distribution of reference impacts and the spatial correlation between reference impacts and the target impact. To illustrate location-based transfer function interpolation and extrapolation, \cref{FIG C7: TF_FSDT_RBF} presents the results based on the calibrated FSDT model and four reference impacts, numbered 12, 14, 22, and 24.

For transfer function interpolation, shown in \cref{FIG C7: TF_FSDT_RBF}(a), two target impacts, 13 and 18, are considered. Target impact 13 is positioned between reference impacts 12 and 14, while target impact 18 is located centrally among all four reference impacts. It can be observed that for both target impacts, the interpolated transfer functions exhibit higher accuracy in the low-frequency range but reduced accuracy at higher frequencies. Additionally, modes within the 2000 Hz to 3000 Hz range are more pronounced for target impact 13 but relatively weaker for target impact 18. This suggests that the RBF method effectively captures and preserves the mode contributions within the interpolation region.

In contrast, \cref{FIG C7: TF_FSDT_RBF}(b) illustrates the transfer function extrapolation for target impacts 1 and 7, both of which lie outside the spatial coverage of the four reference impacts. It is evident that the accuracy of the extrapolated transfer functions cannot be guaranteed across the entire frequency range. Both underestimation and overestimation of peak amplitudes are observed, highlighting the limitations of location-based transfer function extrapolation. These errors arise due to the lack of sufficient reference data beyond the known impact region, leading to a less reliable estimation of structural dynamics in extrapolated areas.

The location-based transfer function estimation using the RBF method highlights the critical role of the spatial distribution of reference impacts in achieving high-precision interpolation and capturing higher-frequency modes. Fortunately, due to the dominance of low-frequency modes in impact dynamics, the RBF method remains effective in estimating transfer functions at low frequencies even with sparse data, leading to accurate impact force deconvolution.

\subsubsection{Constant regularisation (GCV) and adaptive regularisation} \label{subsubsection C7: two regularisation}
For frequency-domain impact force deconvolution, as defined in \cref{EQU: reg deconvolution frequency domain}, the regularisation parameter $\lambda_{reg}$ plays a critical role in suppressing noise amplification caused by uncertainties in sensor signals and transfer function estimates. Traditionally, $\lambda_{reg}$ is treated as a constant, applying uniform regularisation across all frequencies. The optimal constant value is typically selected using the Generalised Cross-Validation (GCV) criterion, as described in \cref{EQU GCV}. However, this global approach often introduces excessive regularisation in the low-frequency range, leading to underestimation of the reconstructed impact force.

\begin{figure}[htb] 
	\centering
		\includegraphics[width=1\columnwidth]{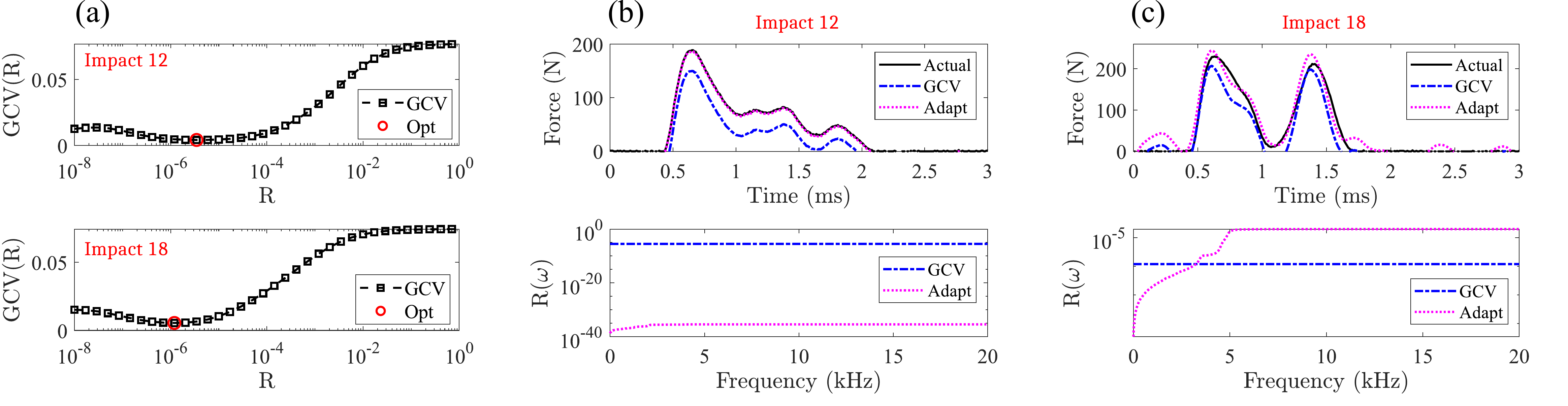}
	\caption{Impact force deconvolution based on RBF interpolated transfer function using four reference impacts, with constant and adaptive regularisation: (a) contant regularisation parameter determined by GCV, (b) force deconvolution for impact 12, (b) force deconvolution for impact 18. }
	\label{FIG C7: force_decon_4RI_12_18}
\end{figure}

\cref{FIG C7: force_decon_4RI_12_18} compares the performance of constant and adaptive regularisation strategies, where the latter is defined based on the FSDT model as in \cref{EQU C7: adaptive R}. The constant regularisation parameters for impacts 12 and 18 are determined using GCV, which identifies optimal values of $3.455 \times 10^{-6}$ and $1.193 \times 10^{-6}$, respectively, as indicated by the red markers in \cref{FIG C7: force_decon_4RI_12_18}(a).

Impact 12 serves as one of the high-fidelity (HF) experimental reference impacts used in FSDT modelling and RBF-based transfer function estimation. For this case, the RBF interpolated transfer function exhibits excellent agreement with both experimental and FSDT predictions in both high- and low-frequency regions. Consequently, the adaptive regularisation $\lambda_{reg}(\omega)$ is virtually negligible—on the order of $10^{-40}$—due to floating-point numerical limits. As shown in \cref{FIG C7: force_decon_4RI_12_18}(b), this minimal regularisation avoids excessive smoothing and leads to improved force reconstruction compared to constant regularisation, which tends to underestimate the peak force due to oversuppression of low-frequency components.

In contrast, impact 18 is in the middle of the four HF reference impacts, allowing for imperfect transfer function interpolation. In this case, the adaptive regularisation $\lambda_{reg}(\omega)$ increases significantly with frequency. Below 4 kHz, the adaptive $\lambda_{reg}(\omega)$ remains smaller than the constant GCV-derived value, while at higher frequencies, it surpasses the constant value, as shown in \cref{FIG C7: force_decon_4RI_12_18}(c). Despite this, the adaptive approach results in a more accurate force estimate overall. It avoids the peak underestimation seen with constant regularisation and slightly overestimates the peak force to approximately 15 N—corresponding to a 6.5$\%$ deviation from the ground truth. This outcome confirms that low-frequency components, which dominate impact dynamics, are especially sensitive to regularisation strength. To minimise underestimation, regularisation should be carefully modulated across frequencies—lighter at low frequencies and stronger at high frequencies. The adaptive scheme achieves this balance and improves the fidelity of impact force reconstruction.

\subsubsection{Impact force reconstruction based on extrapolated transfer functions}
\begin{figure}[htb] 
	\centering
		\includegraphics[width=0.8\columnwidth]{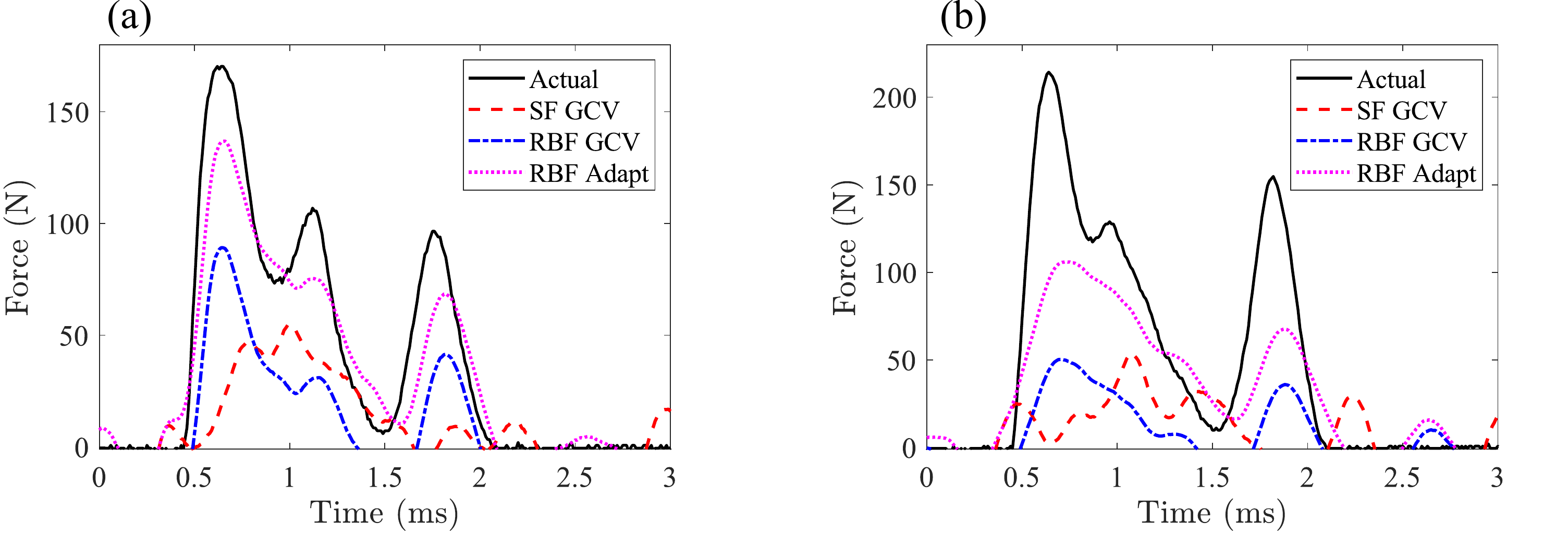}
	\caption{Impact force deconvolution based on extrapolated transfer functions using four reference impacts: (a) impact 7, (a) impact 2. }
	\label{FIG C7: force_decon_4RI_7_2}
\end{figure}

Force reconstruction deteriorates as the target impact moves further from the region covered by the HF reference impacts, transforming the problem into one of transfer function extrapolation. \cref{FIG C7: force_decon_4RI_7_2} illustrates the impact force deconvolution for impacts numbered 7 and 2, based on extrapolated transfer functions using the RBF and SF methods. Since these impacts lie outside the four HF reference impacts, their transfer functions are extrapolated and therefore less accurate, leading to an underestimation of the reconstructed forces compared to the actual forces.

The shape function (SF) method, as detailed in \cref{subapp: shape function}, which extrapolates by adopting the nearest reference transfer function, results in the most inaccurate force reconstruction, often distorting the force waveform. In contrast, the RBF method more accurately preserves the force shape, benefiting from its consistency in transfer function estimation and its 'minimum energy' property.
Furthermore, consistent with \cref{subsubsection C7: two regularisation}, RBF with adaptive regularisation yields a more accurate force estimation compared to constant regularisation. Comparing force reconstructions for the two target impacts, it is evident that greater distance from the reference impacts results in lower reconstructed force magnitudes, due to increased inaccuracies in transfer function extrapolation.

\subsubsection{Impact force reconstruction with increased HF reference impacts}
\label{subsubsection IFR HF variation}
Two strategies are considered to improve force deconvolution accuracy across the structure: (1) increasing the number of HF reference impacts and (2) enhancing the accuracy of the LF model. The first approach is more practical, as increasing the number of HF reference impacts allows for more comprehensive coverage of the structure, facilitating accurate transfer function interpolation. In contrast, improving the LF model’s accuracy to replicate HF experimental data requires detailed structural information—such as precise material properties and boundary conditions—and higher-order modelling techniques, such as FEA. Due to its data-driven nature and boundary condition simplification, the FSDT model alone cannot achieve HF accuracy. 

\begin{figure}[htb] 
	\centering
		\includegraphics[width=0.8\columnwidth]{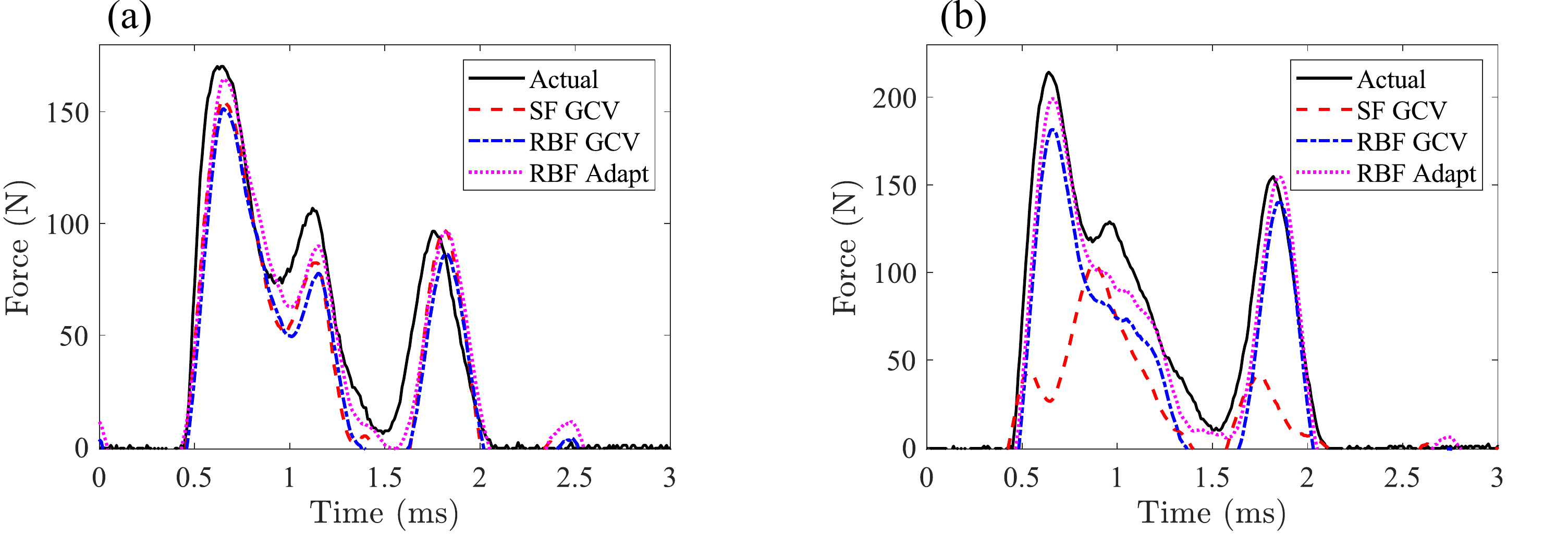}
	\caption{Impact force deconvolution based on interpolated transfer functions using nine reference impacts: (a) impact 7, (b) impact 2. }
	\label{FIG C7: force_decon_9RI_7_2}
\end{figure}

By increasing the number of HF reference impacts from 4 to 9, both impacts 7 and 2 fall within the coverage region of these reference impacts. \cref{FIG C7: force_decon_9RI_7_2} illustrates the force deconvolution results for these two impacts based on interpolated transfer functions using nine reference impacts. Once again, the RBF method demonstrates greater robustness than the SF method in transfer function estimation, consistently capturing the actual force waveform, whereas the SF method fails for impact 2. Additionally, the RBF method exhibits peak force underestimation, though to a lesser extent with adaptive regularisation. Compared to constant regularisation using GCV, adaptive regularisation further mitigates peak force underestimation. For both impacts, the relative peak force error remains within 7$\%$. 

\subsubsection{Impact force uncertainty quantification}
\begin{figure}[htb] 
	\centering
		\includegraphics[width=1\columnwidth]{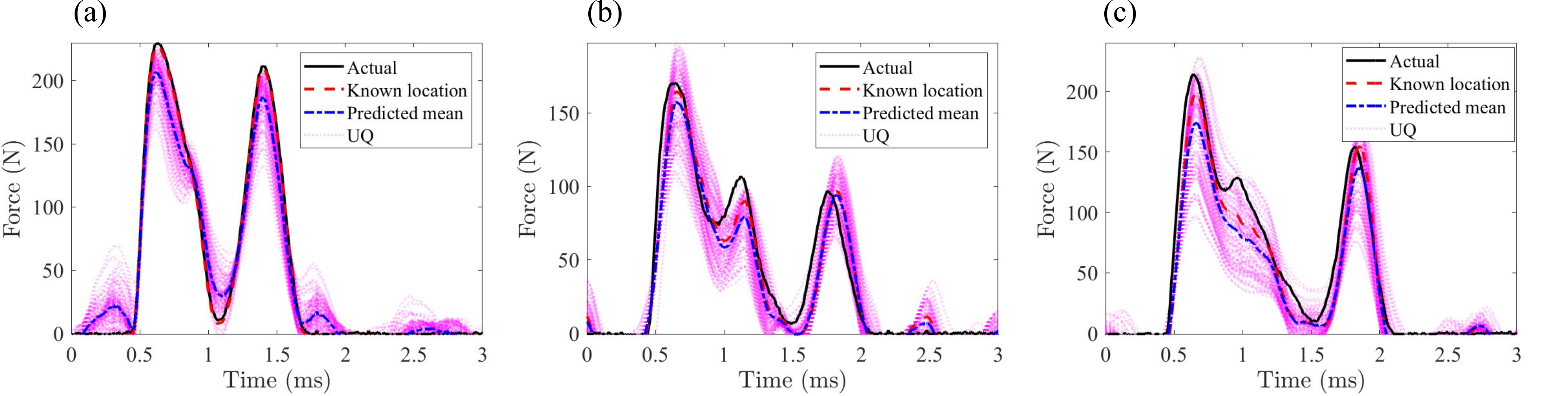}
	\caption{Impact force deconvolution and uncertainty quantification based on localised locations using nine HF reference impacts: (a) impact 18, (b) impact 7, (c) impact 2. }
	\label{FIG C7: force_decon_UQ}
\end{figure}

The preceding analysis of transfer function estimation and impact force deconvolution, presented in \cref{subsubsection C7: two regularisation} to \cref{subsubsection IFR HF variation}, was conducted under the assumption of known impact locations. However, in practical applications, impact localisation is subject to uncertainty. The physics-augmented multi-fidelity GPR framework used for impact localisation provides both a predictive mean and variance, enabling the propagation of localisation uncertainties into force reconstruction. \cref{FIG C7: force_decon_UQ} illustrates the impact force deconvolution and uncertainty quantification for three impacts (numbered 18, 7, and 2), based on GPR-estimated impact location means (with associated variances) and RBF-based transfer function interpolation.

Since transfer function estimation is location-dependent, the discrepancy in force deconvolution between the actual impact location and the GPR-predicted mean location reflects the localisation error. For impacts 18, 7, and 2, these localisation errors are 5.0 mm, 5.5 mm, and 6.9 mm, respectively. The difference between the deconvolved forces at the ‘given location’ (dashed red) and the ‘predicted mean’ (dashed blue) increases from \cref{FIG C7: force_decon_UQ}(a) to (c). Notably, for impact 2, shown in \cref{FIG C7: force_decon_UQ}(c), the predicted mean falls outside the coverage region of the nine HF reference impacts. This results in transfer function extrapolation, which significantly overestimates the transfer function, leading to substantial underestimation of the reconstructed impact force. In contrast, for impacts 18 and 7, where the 95$\%$ confidence interval of the estimated impact location remains within the HF reference coverage, the mean force deconvolution is more accurate.

The uncertainty in deconvolved forces is influenced by the predictive variance of impact localisation. For impacts 18, 7, and 2, the SDs of localisation errors in the x- and y-coordinates are [8.4, 7.7] mm, [8.8, 7.5] mm, and [9.7, 7.5] mm, respectively. The increasing trend in localisation SDs corresponds to the growing uncertainty in the deconvolved forces observed in \cref{FIG C7: force_decon_UQ}.

\section{Conclusions and future work} \label{section Conclusions and future work}
This study presents a sparse data-driven framework for impact localisation and force reconstruction in composite aerostructures, underpinned by a First-Order Shear Deformation Theory (FSDT) plate model that is constructed entirely from observed structural responses. By extracting material properties via dispersion relations and boundary conditions through modal analysis, the approach bypasses the need for prior structural knowledge and leverages physics-augmented multi-fidelity GPR for impact localisation. The constructed model also captures mode shape complexity, which informs an adaptive regularisation strategy for impact force deconvolution. Furthermore, the framework systematically propagates uncertainties from the localisation stage into the force reconstruction process, providing probabilistic confidence bounds for estimated forces.

Experimental validation on a CFRP composite plate subjected to low-velocity impact testing demonstrates the method's effectiveness in real-world scenarios, particularly in settings with sparse training data and limited sensor coverage. The key conclusions are summarised as follows:
\begin{itemize}
    \item \textbf{Sparse data dependence for impact localisation:} Experimental results show that only four reference impacts and four passive sensors are sufficient to infer effective material properties—through GVPs—and to generate augmented training data for accurate GPR-based localisation, including in extrapolated regions beyond the reference set.
    \item \textbf{Data dependence in location-based transfer function interpolation:} While high-frequency modal responses exhibit significant spatial variation requiring denser sampling for accurate reconstruction, low-frequency modes dominate the impact dynamics. This enables robust impact force estimation from sparsely distributed training data, making the method well-suited for in-situ monitoring applications.

    \item \textbf{FSDT-based model construction from sparse observations:} The use of a physics-based FSDT model, guided by dispersion and modal analysis, ensures that critical structural dynamics are preserved even under limited data availability, offering a physically interpretable and scalable approach to modelling composite structures.
    
    \item \textbf{Interpolation vs. extrapolation in location-based transfer function estimation:} The study highlights that while interpolation performs reliably for nearby locations, extrapolation can introduce greater uncertainty. This underpins the necessity of probabilistic modelling and uncertainty quantification in sparse scenarios.
\end{itemize}

By integrating data-driven FSDT modelling, modal analysis, and machine learning, the proposed approach offers a computationally efficient and generalisable framework for impact identification and uncertainty quantification. This significantly reduces experimental effort while maintaining high-fidelity impact force estimation, making it well-suited for real-world applications.

Several avenues are available to extend this research: 
\begin{itemize}
    \item Non-parametric kernel design and physics-informed GPR: Further development of composite or non-parametric kernels guided by physical constraints (e.g., wavefront geometry or dispersion) could enhance GPR accuracy and robustness under varying environmental and operational conditions.
    \item Multi-scale modelling and transfer learning: To improve scalability and applicability across different structural geometries, transfer learning methods could be explored to transfer the learned physics and data representations from laboratory-scale specimens to full-scale aerostructures.
    \item Integration with digital twins: Embedding the proposed method within a digital twin framework would enable real-time structural state awareness, continuous health monitoring, and in-situ model updates based on operational data streams.
\end{itemize}

\appendix
\section{Appendix: Equations of motion for symmetrically-laminated FSDT plate} \label{Appendix A}
In a symmetrically laminated FSDT plate, the bending-extensional coupling stiffness $\mathbf{B}$ and the coupling inertial $I_2$ vanish, resulting in a complete decoupling of in-plane and out-of-plane motions \cite{abrate_impact_1998}. Ignoring the external force, these governing equations can be expressed using a differential operator as $\mathcal{L} \mathbf{d} = 0, \quad \mathbf{d} = [u_0, \; v_0, \; w_0, \; \psi_x, \; \psi_y]^T$, where the operators $\mathcal{L}$ are given as:
\begin{equation}
\begin{aligned} \label{EQU appc: equations of motion FSDT differential operator symmetric}
&L_{11} =  A_{11} \frac{\partial ^2 }{\partial x^2} + 2 A_{16} \frac{\partial ^2 }{\partial x \partial y} + A_{66} \frac{\partial ^2 }{\partial y^2} - I_1\frac{\partial ^2 }{\partial t^2}, \;
L_{12} = A_{16} \frac{\partial ^2 }{\partial x^2} +  (A_{12} + A_{66}) \frac{\partial ^2 }{\partial x \partial y} +  A_{26} \frac{\partial ^2 }{\partial y^2}, \\
&L_{13}  =  0, \;
L_{14} = 0, \;
L_{15} = 0, \\
&L_{22} = A_{66} \frac{\partial ^2 }{\partial x^2} + 2 A_{26} \frac{\partial ^2 }{\partial x \partial y} + A_{22}  \frac{\partial ^2 }{\partial y^2}  - I_1\frac{\partial ^2 }{\partial t^2}, \\ 
&L_{23} = 0, \;
L_{24} = 0, \; 
L_{25} =  0, \\
&L_{33} =  (A_{55} \frac{\partial^2  }{\partial x^2} + 2A_{45} \frac{\partial^2  }{\partial x \partial y} + A_{44} \frac{\partial^2  }{\partial y^2}) -  I_1\frac{\partial ^2 }{\partial t^2}, \;
L_{34} = A_{55}\frac{\partial }{\partial x} + A_{45}\frac{\partial }{\partial y}, \;
L_{35} = A_{45}\frac{\partial }{\partial x} + A_{44}\frac{\partial }{\partial y}, \\
&L_{44} = D_{11}\frac{\partial^2 }{\partial x^2} + 2D_{16}\frac{\partial^2 }{\partial x\partial y} + D_{66}\frac{\partial^2 }{\partial y^2} -A_{55} - I_3\frac{\partial ^2 }{\partial t^2}, \\
&L_{45} = D_{16}\frac{\partial^2 }{\partial x^2} + (D_{12}+D_{66}) \frac{\partial^2 }{\partial x\partial y} + D_{26}\frac{\partial^2 }{\partial y^2} -A_{45}, \\ 
&L_{55} = D_{66}\frac{\partial^2 }{\partial x^2} + 2D_{26} \frac{\partial^2 }{\partial x\partial y} + D_{22}\frac{\partial^2 }{\partial y^2} -A_{44} - I_3\frac{\partial ^2 }{\partial t^2}. 
\end{aligned}
\end{equation}
This decoupling is reflected in the governing equations, where $L_{1j} = L_{2j}= L_{j1} = L_{j2}=0$ for $j = 3,4,5$, ensuring independent in-plane and out-of-plane dynamics.

\section{Appendix: Dispersion relation derived from FSDT plate} \label{Appendix B}
For FSDT, the $3 \times 3$ matrix $\mathbf{G}_o$ is given by \cite{abrate_impact_1998, moon_theoretical_1973, tan_wave_1982, tang_lamb_1988, chow_propagation_1971, jeong_wavelet_2000}:
\begin{equation}
\begin{aligned} \label{EQU C2: G matrix flexural wave FSDT}
\mathbf{G}_o = \left[ \begin{matrix}
g_{11} & g_{12} & g_{13}  \\
g_{12} & g_{22} & g_{23}  \\
g_{13} & g_{23} & g_{33}
\end{matrix} \right], \; \mathbf{G}_o \left[ \begin{matrix}
W \\
\Phi_x \\
\Phi_y
\end{matrix} \right] = 0
\end{aligned}
\end{equation}
The elements $g_{ij}$, for $i,j = 1,2,3$, of the coefficient matrix $\mathbf{G}_0$ are given by:
\begin{equation}
\begin{aligned} \label{EQU C2: g_ij flexural wave FSDT}
&g_{11} = -(A_{44}\sin^2 \theta + A_{55}\cos^2 \theta + 2A_{45}\sin\theta \cos\theta)k^2 +I_{1}\omega^2, \\
&g_{12} = i(A_{55}\cos \theta+A_{45}\sin \theta)k, \;
g_{13} = i(A_{45}\cos \theta+A_{44}\sin \theta)k, \\ 
&g_{22} = (D_{11}\cos^2\theta+2D_{16}\sin\theta \cos\theta+D_{66}\sin^2\theta)k^2 +A_{55}-I_3\omega^2, \\ 
&g_{23} = [(D_{12}+D_{66})\sin\theta\cos\theta+D_{16}\cos^2\theta + D_{26}\sin^2\theta] k^2 + A_{45}, \\ 
&g_{33} = (D_{22}\sin^2\theta + D_{66}\cos^2\theta+2D_{26}\sin\theta\cos\theta)k^2 + A_{44}-I_3\omega^2, 
\end{aligned}
\end{equation}
where $i$ denotes the imaginary unit ($i^2 = -1$). For a non-trivial solution, the determinant of the matrix $\mathbf{G}_o$ must be zero:
\begin{equation}
\begin{aligned} \label{EQU C2: dispersion equation flexural wave FSDT}
\det \mathbf{G}_o = 0, 
\end{aligned}
\end{equation}
leading to a sextic polynomial equation in terms of the wave frequency $\omega$, which contains only even-degree terms. For each wavenumber $k$, there are three positive frequency solutions to this polynomial equation, corresponding to three values of phase velocity $ v_p = \omega / k$ and three wave modes A0, A1, and A2.

\section{Appendix: Modal analysis by Rayleigh-Ritz variational method} \label{Appendix C}
The mass matrix M for modal analysis using the Rayleigh–Ritz method for the symmetrically-laminated FSDT plate, as defined in \cref{EQU C7: Rayleigh-Ritz}, is given by \cite{ni_aeroelastic_2023}:
\begin{equation}
\begin{aligned} 
\mathbf{M} = \left[ \begin{matrix}
\mathbf{M}_1 & 0 & 0 & 0 & 0 \\
0 & \mathbf{M}_1 & 0 & 0 & 0 \\
0 & 0 & \mathbf{M}_1 & 0 &  0 \\
0 & 0 & 0 & \mathbf{M}_3 & 0 \\
0 & 0 & 0 & 0 & \mathbf{M}_3
\end{matrix} \right], 
\end{aligned} 
\end{equation}
where $\mathbf{M}_i = \int_{\Omega} I_i\mathbf{B}^T\mathbf{B} d \Omega$, $\mathbf{B}$ is the row vector of basis functions, expressed in \cref{EQU C7: displacement field}.  

The detailed expression of the stiffness matrix $\mathbf{K}_p$ from plate structure is given by:
\begin{equation}
\begin{aligned} 
\mathbf{K}_p = \left[ \begin{matrix}
\mathbf{K}_{11} & \mathbf{K}_{12}& 0 & 0 & 0 \\
\mathbf{K}_{12}^T & \mathbf{K}_{22} & 0 & 0 & 0 \\
0 & 0 & \mathbf{K}_{33}  & \mathbf{K}_{34} &  \mathbf{K}_{35} \\
0 & 0 & \mathbf{K}_{34}^T & \mathbf{K}_{44} & \mathbf{K}_{45} \\
0 & 0 & \mathbf{K}_{35}^T & \mathbf{K}_{45}^T & \mathbf{K}_{55} \\
\end{matrix} \right],  \; 
\mathbf{K}_{ij} = \int_{\Omega} \left[\mathbf{B}^T \; (\frac{\partial \mathbf{B}}{\partial x})^T \;  (\frac{\partial \mathbf{B}}{\partial y})^T \right] \mathbf{C}_{ij} \left[ \begin{matrix} \mathbf{B} \\ \frac{\partial \mathbf{B}}{\partial x} \\  \frac{\partial \mathbf{B}}{\partial y} \end{matrix} \right] d \Omega, 
\end{aligned} 
\end{equation}
where $\mathbf{C}_{ij}$ are 3-by-3 matrix consisting of plate-level stiffness:
\begin{equation}
\begin{aligned} 
&\mathbf{C}_{11} = \left[ \begin{matrix} 
0 & 0 & 0 \\
0 & A_{11} & A_{16} \\
0 & A_{16} & A_{66} \\
\end{matrix} \right], \; 
\mathbf{C}_{12} = \left[ \begin{matrix} 
0 & 0 & 0 \\
0 & A_{16} & A_{12} \\
0 & A_{66} & A_{26} \\
\end{matrix} \right], \; 
\mathbf{C}_{22} = \left[ \begin{matrix} 
0 & 0 & 0 \\
0 & A_{66} & A_{26} \\
0 & A_{26} & A_{22} \\
\end{matrix} \right], \\ 
&\mathbf{C}_{33} = \left[ \begin{matrix} 
0 & 0 & 0 \\
0 & A_{55} & A_{45} \\
0 & A_{45} & A_{44} \\
\end{matrix} \right], \; 
\mathbf{C}_{34} = \left[ \begin{matrix} 
0 & 0 & 0 \\
A_{55} & 0 & 0 \\
A_{45} & 0 & 0 \\
\end{matrix} \right], \; 
\mathbf{C}_{35} = \left[ \begin{matrix} 
0 & 0 & 0 \\
A_{45} & 0 & 0 \\
A_{44} & 0 & 0 \\
\end{matrix} \right], \\ 
&\mathbf{C}_{44} = \left[ \begin{matrix} 
A_{55} & 0 & 0 \\
0 & D_{11} & D_{16} \\
0 & D_{16} & D_{66} \\
\end{matrix} \right], \; 
\mathbf{C}_{45} = \left[ \begin{matrix} 
A_{45} & 0 & 0 \\
0 & D_{16} & D_{12} \\
0 & D_{66} & D_{26} \\
\end{matrix} \right], \;  
\mathbf{C}_{55} = \left[ \begin{matrix} 
A_{44} & 0 & 0 \\
0 & D_{66} & D_{26} \\
0 & D_{26} & D_{22} \\
\end{matrix} \right].
\end{aligned} 
\end{equation}

The detailed expression of the stiffness matrix $\mathbf{K}_b$ from boundary conditions is given by:
\begin{equation}
\begin{aligned} 
\mathbf{M}_b = \int_{\Gamma} \mathbf{B}^T \diag(k_t^u\mathbf{I}, k_t^v\mathbf{I}, k_t^w\mathbf{I}, k_r^x\mathbf{I}, k_r^y \mathbf{I})\mathbf{B} d\Gamma,  
\end{aligned} 
\end{equation}
where $[k_t^u, k_t^v, k_t^w, k_r^x, k_r^y]$ are three translational stiffness and two rotational stiffness of the artificial springs and acting on boundaries $\Gamma$, $\mathbf{I}$ is the identity matrix with the size of the number of basis functions in $\mathbf{B}$.

\section{Appendix: Data-driven, location-based transfer function estimation} \label{section transfer function estimation}
The reference transfer functions $\hat{H}$ in frequency domain between the reference impact forces $F(t)$ at known impact locationss $Q$ and the sensor signal $S(t)$ measured at sensor locatios $L$ can be directly estimated using Welch's averaged periodogram method \cite{stoica_spectral_2005}:
\begin{equation}
\begin{aligned}
\hat{H}(\omega; Q, L) = \frac{P_{fs}(\omega; Q, L) }{P_{ff}(\omega; Q, L)}, 
\end{aligned}
\end{equation}
where $P_{fs}(\omega; Q, L)$ represents the cross power spectral density between the impact force and the sensor signal, and $P_{ff}(\omega; Q, L)$ denotes the power spectral density of the impact force itself. Data-driven transfer function estimation aims to infer the transfer function $\hat{h}(\omega; q, L)$ at location $q$ based on known transfer functions $\hat{H}(\omega; Q, L)$ at reference locations $Q$.

\subsection{Shape Function (SF) method}
\label{subapp: shape function}
The Shape Function (SF) method \cite{park_monitoring_2009, chen_impact_2012, xu_determination_2016} estimates the transfer function at an impact location 
$q$ by interpolating based on the transfer functions at four surrounding reference impact locations. As depicted in \cref{FIG C7: SF method}, this approach involves mapping an irregular quadrilateral cell in the physical domain onto a standardised parametric space with a non-dimensional coordinate system. In this parametric space, the transformed element maintains the same topology as the original, with nodal coordinates represented by $\xi, \eta$. The transfer function at $q$ is then approximated as a weighted sum of the transfer functions at the four reference locations:
\begin{figure}[htb] 
	\centering
		\includegraphics[width=0.6\columnwidth]{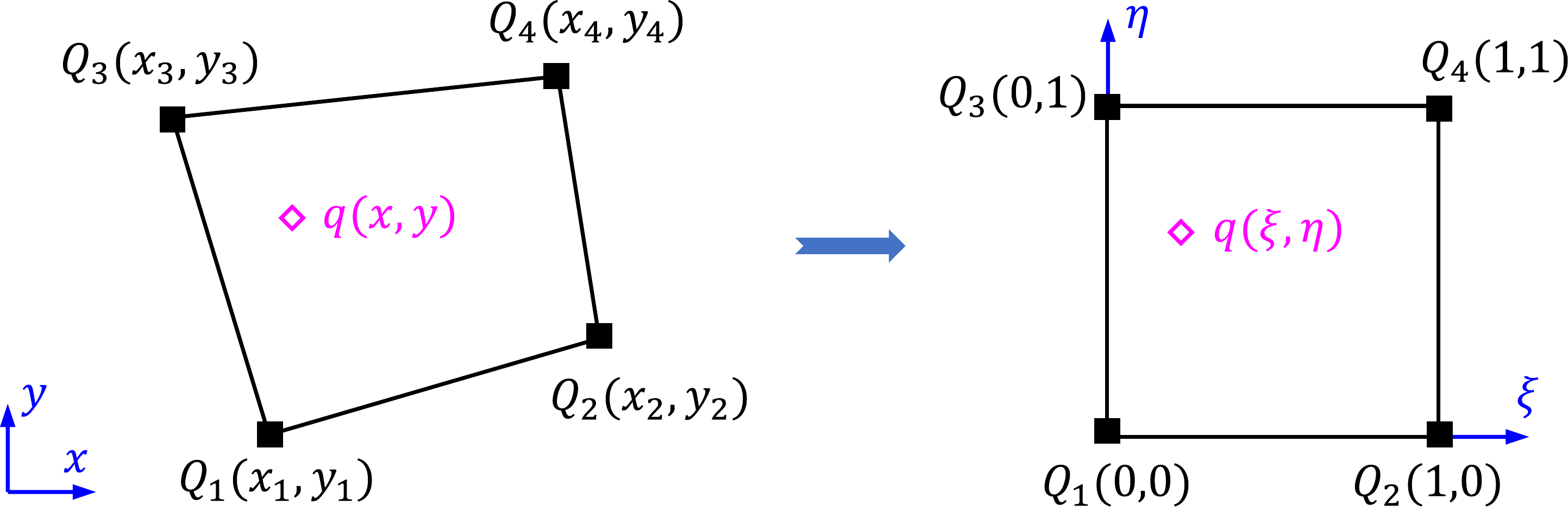}
	\caption{Shape Function (SF) method for transfer function interpolation.}
	\label{FIG C7: SF method}
\end{figure}

\begin{equation}
\begin{aligned} \label{EQU: SF interpolant}
h(x, y) = \sum_{i=1}^{4}\lambda_i h_i(x_i, y_i)
\end{aligned}
\end{equation} 
where $h_i(x_i, y_i)$ denotes the transfer function at the i-th reference location, and $\lambda_i$ are the shape functions that establish the relationship between the physical coordinates (x,y) and the parametric coordinates ($\xi, \eta$). For bilinear interpolation in two dimensions, these shape functions are defined as:
\begin{equation}
\begin{aligned} \label{EQU: SF shape functions}
\lambda_1(\xi, \eta) = (1-\xi)(1-\eta), \; \lambda_2(\xi, \eta) = \xi(1-\eta), \; 
\lambda_3(\xi, \eta) = (1-\xi)\eta, \; \lambda_4(\xi, \eta) = \xi \eta, 
\end{aligned}
\end{equation} 
where the local coordinates are determined by: $\xi = \frac{x-x_1}{x_2-x_1}$, $\eta= \frac{y-y_1}{y_3-y_1}$. The accuracy of the SF method relies on the proximity of the reference impact locations to the target impact position. While it provides a localised and computationally efficient interpolation, its effectiveness diminishes as the target impact location moves further from the reference points. This limitation arises because interpolation errors increase with distance, making the SF method less suitable for scenarios where the reference impact data are sparse or when the target impact falls outside the convex hull of reference points.

\subsection{Radial Basis Function (RBF) method} \label{subapp: RBF}
Radial basis functions (RBFs) are widely employed in function approximation, data interpolation, and machine learning due to their ability to model complex relationships from scattered data points. They are particularly well-suited for multi-dimensional interpolation, as they do not require a structured grid and can effectively capture variations in the underlying function with high accuracy. The general principle of the RBF method is that for a given set of $N$ data points $\mathbf{x}_i, i =1,2,...,N$ and corresponding data values $h_i, i =1,2,...,N$, a set of basis functions $\psi(\left|\mathbf{x}-\mathbf{x}_i\right|)$ centred at given data points is chosen such that a linear combination of these functions satisfies the interpolation conditions:
\begin{equation}
\begin{aligned} \label{EQU: RBF interpolant}
h(\mathbf{x}) = \sum_{i=1}^{N} \lambda_i\psi(\left|\mathbf{x} - \mathbf{x}_i \right|) = \sum_{i=1}^{N} \lambda_i\psi(r_i), 
\end{aligned}
\end{equation} 
where $g(\mathbf{x})$ is the interpolated function, $\psi(r), r\ge 0$ is the radial basis function, $\lambda_i$ are the coefficients to be determined, $\mathbf{x}$ is the location of interest, $\left| \cdot \right|$ denotes the abs operator. The expansion coefficients are determined from the interpolation conditions $h(\mathbf{x}_i) = h_i, i =1,2,...,N$, leading to the following linear system:
\begin{equation}
\begin{aligned}
\boldsymbol{\Psi} \boldsymbol{\lambda}= \mathbf{h}, 
\end{aligned}
\end{equation} 
where $\boldsymbol{\Psi}$ is $N$ by $N$ matrix, whose $i, j$-th element is given by $\psi(\left|\mathbf{x}_i-\mathbf{x}_j\right|)$.

Common choices of radial basis functions include Gaussian, multiquadric, and thin-plate splines (TPS), with the selection dependent on smoothness and computational efficiency requirements. To enhance stability, it is recommended to incorporate additional polynomial terms into the RBF interpolant:
\begin{equation}
\begin{aligned} \label{EQU: AURBF interpolant}
h(\mathbf{x}) = \sum_{i=1}^{N} \lambda_i\psi(\left|\mathbf{x} - \mathbf{x}_i \right|) + \sum_{j=1}^{M} \varsigma_j p_j(\mathbf{x}), 
\end{aligned}
\end{equation} 
where $p_j(\mathbf{x}), j =1,2,...,M$ form the basis for the polynomial space $P_l(\mathbb{R}^{d_{i}})$, which consists of $d_i$-variate polynomials of maximum degree $l_p$. The dimension $M$ of the polynomial space is given by $M = C^{d_i+l_p}_{d_i}$. Additional constraints ensure the interpolation system remains solvable:
\begin{equation}
\begin{aligned} \label{EQU: AURBF interpolant constrains}
\sum_{i=1}^{N} \lambda_i p_j(\mathbf{x}_i) = 0, \;  j=1,2,...,M.
\end{aligned}
\end{equation} 
Combining these equations, the coefficients $\lambda_i$ and $\varsigma_j$ for the augmented RBF interpolation are determined by solving:
\begin{equation}
\begin{aligned}
\left[\begin{matrix}
\boldsymbol{\Psi} & \mathbf{P} \\
\mathbf{P}^T & \mathbf{0}
\end{matrix} \right] \left[\begin{matrix}
\boldsymbol{\lambda} \\
\boldsymbol{\varsigma} 
\end{matrix} \right] = \left[\begin{matrix}
\mathbf{h} \\
\mathbf{0}
\end{matrix} \right].
\end{aligned}
\end{equation} 

Consequently, the RBF method is highly beneficial for data-driven transfer function estimation, where the impact location serves as the input, provided that the transfer functions at reference impact locations are known. Additionally, the computational requirements are relatively low, as the method involves only small-scale matrix inversion, with the matrix size equal to (or slightly larger than) the number of reference impact locations. Simone et al. \cite{simone_hierarchical_2019} has demonstrated that thin-plate spline (TPS) RBF ($\psi(r) = r^2 \ln r$) is particularly suitable for this application due to its smoothness and ability to minimise the so-called ‘bending energy’ \cite{bookstein_principal_1989}. Moreover, TPS requires no hyperparameter tuning, thereby reducing computational overhead. For a two-dimensional structural space ($d_i=2$), Micchelli’s theorem \cite{micchelli_interpolation_1984} guarantees that the augmented TPS RBF method, as expressed in \cref{EQU: AURBF interpolant}, is uniquely solvable when $l_p=1$.

\section*{CRediT authorship contribution statement}
\textbf{Dong Xiao}: Conceptualization, Methodology, Software, Data curation, Formal analysis, Writing - Original draft preparation, Writing - Review Editing. \textbf{Zahra Sharif-Khodaei}: Conceptualization, Methodology, Writing - Review Editing, Supervision. \textbf{M. H. Aliabadi}: Conceptualization, Methodology, Writing - Review Editing, Supervision.

\section*{Declaration of competing interest}
The authors declare that they have no known competing financial interests or personal relationships that could have appeared to influence the work reported in this paper.

\section*{Acknowledgements}
The first author acknowledges and expresses gratitude to the China Scholarship Council for providing a scholarship (No. [2021]339) to fund his Ph.D. studies.

\section*{Data availability}
Data will be made available on request.

\small{

}

\end{document}